\documentclass{article}
\usepackage{cite}
\usepackage{amsmath,amssymb,amsfonts}
\usepackage{algorithmic}
\usepackage{hyperref}
\usepackage{graphicx}
\usepackage{subcaption}
\usepackage{textcomp}
\usepackage{xcolor}
\usepackage{cleveref}
\usepackage{float}
\def\BibTeX{{\rm B\kern-.05em{\sc i\kern-.025em b}\kern-.08em
    T\kern-.1667em\lower.7ex\hbox{E}\kern-.125emX}}

\usepackage[letterpaper, margin=0.75in]{geometry}
    
\begin{document}

\title{Radio Frequency Amplitude-Modulation to Frequency-Modulation Signal Converter}
\author{Rishab Parthasarathy\footnote{equal contribution} \and Michael Popik$^{*}$ \and Noah Haefner$^{*}$}

\date{}

\maketitle

\begin{abstract}

In this project, we wanted to discover an analog topology that could effectively convert amplitude-modulated (AM) signals to frequency-modulated (FM) signals, while also ensuring that both sets of signals were within their respective radio frequency (RF) bands. To that end, an effective topology for doing so was developed, characterized, and demonstrated, requiring the ability to de-modulate incoming signals from the AM radio band---spanning from 530 kHz to 1700 kHz \cite{fcc_am_query}---and re-modulate these signals into the FM radio band---spanning from 88 MHz to 108 MHz \cite{fcc_fm_radio}. These bands are separated by roughly 86 MHz, presenting the need for the topology to radically alter the incoming frequency before re-broadcasting. At its simplest implementation, this required an AM demodulation circuit coupled to a voltage controlled oscillator (VCO). Together, these two circuits translated variations in the incoming envelope signal to variations in the output frequency while still maintaining high-fidelity audio, similar to how existing radio receiving and broadcasting are done. Altogether, the project not only developed a working system but also provided valuable instruction in the design, analysis, and construction of effective RF circuits---invaluable to future endeavors within analog electronics.

\end{abstract}

\section{Background and Motivation}

Before frequency modulation (FM) radio receivers became common, FM-to-AM converters allowed those with older amplitude modulation (AM) radios to receive FM broadcasts. However, now that FM dominates, AM radios are becoming less common, so the inverse problem applies—--it is easier to receive FM signals than AM signals. So, to preserve access to an entire section of the radio band, it is now necessary to do AM-to-FM conversion. 
\newline
\newline
First, before exploring conversion, we provide a short theoretical explanation of AM and FM. In both cases, to transmit signals over long distances while limiting attenuation, the signal is modulated by combining it with a higher-frequency carrier \cite{byjus_amplitude_modulation}. Initially, modulation was amplitude-modulated: the carrier frequency is invariant, and the lower frequency envelope carries the signal, encoding the signal as variations in the envelope amplitude. An example of amplitude modulated can be seen in Figure~\ref{fig:AM_FM} with an 800 kHz carrier, a 1 kHz signal, and a modulation depth of 0.75. This situation can be mathematically described by \cite{arar_am_modulation}:

\begin{align*}
x_m(t) &= \cos(2\pi \cdot 1000\cdot t) - \text{Signal}\\
x_c(t) &= \cos(2\pi \cdot 800\cdot10^3\cdot{}t) - \text{Carrier Wave}\\
x_{mod}(t) &= (x_m(t) + 0.75) \cdot A\cos(2\pi \cdot 800\cdot10^3\cdot{}t) - \text{AM Modulated Output} \\
\end{align*}

\begin{raggedright}
However, amplitude modulation suffers from low signal fidelity, making it a poor choice for transmitting signals with a wide range of frequencies \cite{byjus_amplitude_modulation}. So, engineers invented frequency modulation: the amplitude is now invariant, and the frequency varies to encode information. FM offers better fidelity than AM, although over shorter distances, but nonetheless, FM radios now dominate. In Figure~\ref{fig:AM_FM} is an example of FM with a 10 kHz carrier, a 100 Hz signal, and a modulation sensitivity of $10^{10}$. It should be noted that this is a contrived example designed to produce a plot where the reader can observe the change in the wave frequency with time. In FM radio, the actual modulation frequencies are much higher, but this makes the effect more difficult to view in time. This situation can be mathematically described by \cite{tretter_ch8_fm}:
\end{raggedright}

\begin{align*}
x_m(t) &= \cos(2\pi \cdot 100\cdot t) - \text{Signal}\\
x_c(t) &= \cos(2\pi \cdot10^4\cdot{}t) - \text{Carrier Wave} \\
x_{mod}(t) &= \cos(2\pi \cdot 10^4 \cdot t + 10^{10} \int_0^t x_m(\tau) d\tau) - \text{FM Modulated Output}
\end{align*}

Example figures of both AM and FM for the reader to have a visual sense of these processes are given below:

\begin{figure}[h!]
  \centering
  \begin{subfigure}[b]{0.45\textwidth}
    \includegraphics[width=\linewidth]{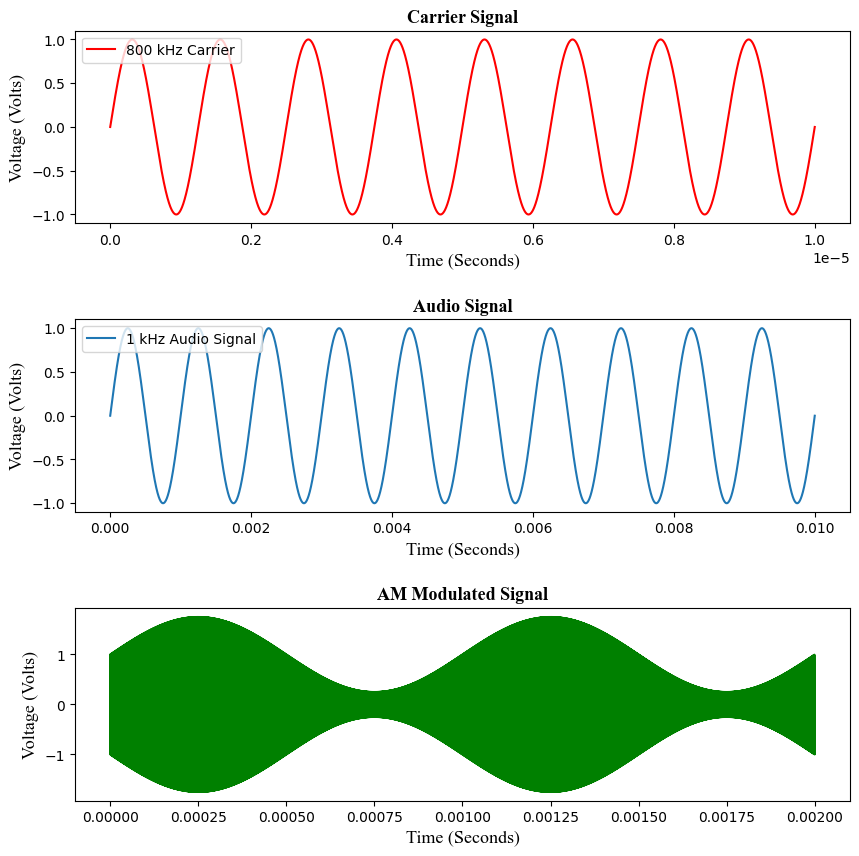}
    \caption{AM Modulation Example}
    \label{fig:AM}
  \end{subfigure}
  \hfill
  \begin{subfigure}[b]{0.45\textwidth}
    \centering
    \includegraphics[width=\linewidth]{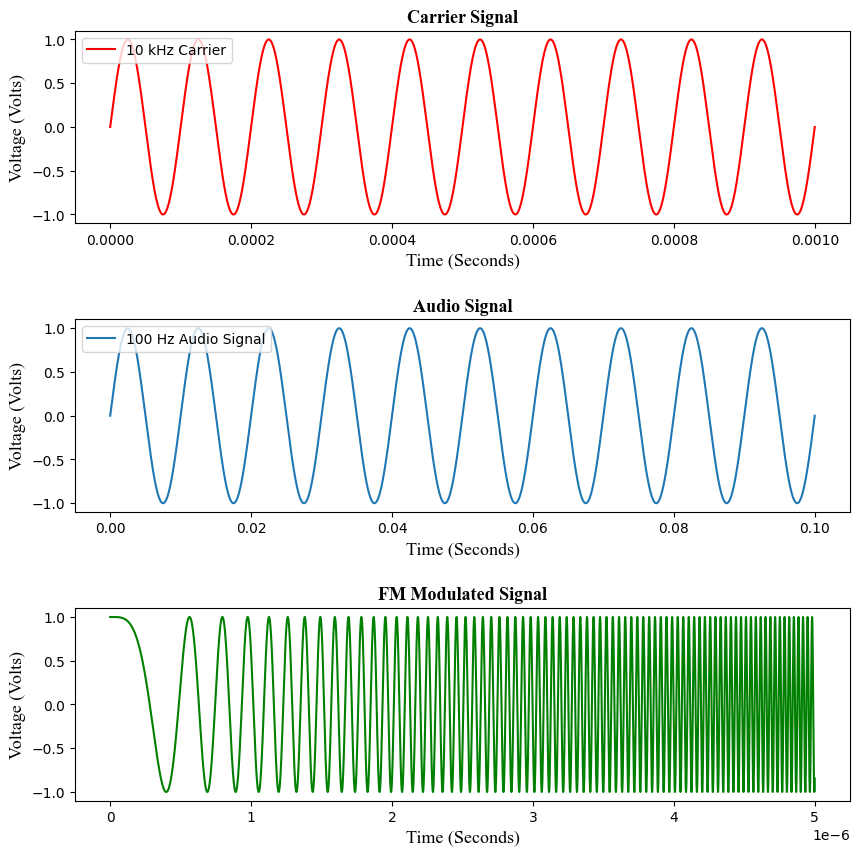}
    \caption{FM Modulation Example}
    \label{fig:FM}
  \end{subfigure}
  \caption{Visualization of AM and FM in the Time Domain}
  \label{fig:AM_FM}
\end{figure}

\begin{raggedright}
For RF applications, the carrier frequencies for AM are between 530 and 1700 kHz \cite{fcc_am_query}, and the carrier frequencies for FM are between 88 and 108 MHz \cite{fcc_fm_radio}---aligning with our contrived AM modulation example but deviating from our contrived FM modulation example.
\end{raggedright}
\newline
\newline
To reiterate Figure~\ref{fig:AM_FM}, the AM signal has constant frequency but varying amplitude, and the FM signal has constant amplitude but varying frequency. The challenge lies in preserving the information in the envelope of the AM wave while converting it into the varying frequencies of the FM wave, without losing any information in the process.
\newline
\newline
Finally, although the intermediate frequency (IF) does not play a major role in our project, it will still be discussed in terms of future improvements. So, a brief introduction here is justified. Essentially, the IF, which is 455 kHz for AM, allowed the expansion of radio by cheapening receivers \cite{efy_intermediate_frequency}. In cheaper receivers, all that is necessary is for a local oscillator frequency to mix with the incoming broadcast frequency, producing two frequencies: $f_b + f_o$ and $f_b - f_o$, where $f_b$ is the broadcast frequency and $f_o$ the local oscillator frequency \cite{efy_intermediate_frequency}. Very much on purpose, $f_b - f_o = f_i$, where $f_i = 455$ kHz---the IF. That way, all receivers can be optimized for the IF instead of requiring a robust frequency response over the wide AM band \cite{efy_intermediate_frequency}. 
\newline
\newline
For our motivation, we sought to build an AM-to-FM converter that would allow us to capture more “old-fashioned” AM signals on “modern” FM devices. Much to our excitement, this project encompassed a full range of radio frequency (RF) electrical engineering: receiving, converting, and transmitting RF signals. We expected that this project would present a high level of difficulty and embraced the challenge. Our goal was to produce a system that effectively takes an AM signal and can convert this to FM audio without distortion or clipping, producing a disappointment free experience for any user.

\section{System Design}

As with most engineering processes, it was helpful to first define non-technical goals for the final product and then use these goals to inform technical design decisions. So, for instance, since we sought to reduce any clipping or distortion during the transition from AM to FM, then it was necessary to leverage topologies that did not degrade the signal-to-noise ratio too much or exceed power supply constraints. And, since we wanted to both receive in the AM band and retransmit in the FM band, then it was necessary to leverage topologies that functioned at RF. Already, leveraging this strategy greatly simplifies the design process by quickly eliminating unviable solutions.
\newline
\newline
Since our end goal was to hear high-fidelity audio at both ends of our system, it was necessary to implement circuit topologies for the two main functions to achieve that goal: faithful demodulation and re-modulation. We must demodulate incoming AM signals to audio frequencies and then re-modulate these audio frequencies back to RF FM signals. We can define our system goals in a semi-technical manner:

\begin{enumerate}
    \item Faithfully demodulate signals within the AM band (530 kHz - 1.7 MHz) $\rightarrow$ Requires high-fidelity demodulation topology
    \item Faithfully re-modulate signals back into the FM band (88 MHz - 108 MHz)  $\rightarrow$ Requires high-fidelity topology that       oscillates at different frequencies based on an incoming signal
\end{enumerate}

\begin{raggedright}
These two goals were accomplished using an AM demodulation topology coupled to a local voltage-controlled oscillator (VCO) topology. For AM demodulation, this was completed asynchronously---independent of the phase shift between the carrier wave and the IF---using an improved envelope detector composed of an op-amp-based full-wave rectifier and an RC low-pass filter. Then, the demodulated signal was passed to a voltage-controlled oscillator (VCO), which altered its output frequency based on the input voltage. We implemented a VCO using a Colpitts oscillator topology built on perfboard, achieving frequencies within the FM band from 66 to 102 MHz. With proper attenuation and buffering, the output could be retransmitted as FM radio, as measured by a software-defined radio (SDR). Overall, our completed system allows a user to listen to a particular AM frequency, convert it to FM, and retransmit it, allowing anyone to listen to their favorite AM radio stations on an FM radio. In addition, from testing, the system provides output audio at a relatively high fidelity to the input. Achieving this, both of our goals are fulfilled.
\end{raggedright}

\subsection{Top Level System Schematic}

At its core, our project can be divided into three main blocks:

\begin{enumerate}
    \item AM Demodulation
    \item AM-to-FM Conversion
    \item Output Buffering
\end{enumerate}

\begin{figure*}[h!]
    \centering
    \includegraphics[width=0.6\linewidth]{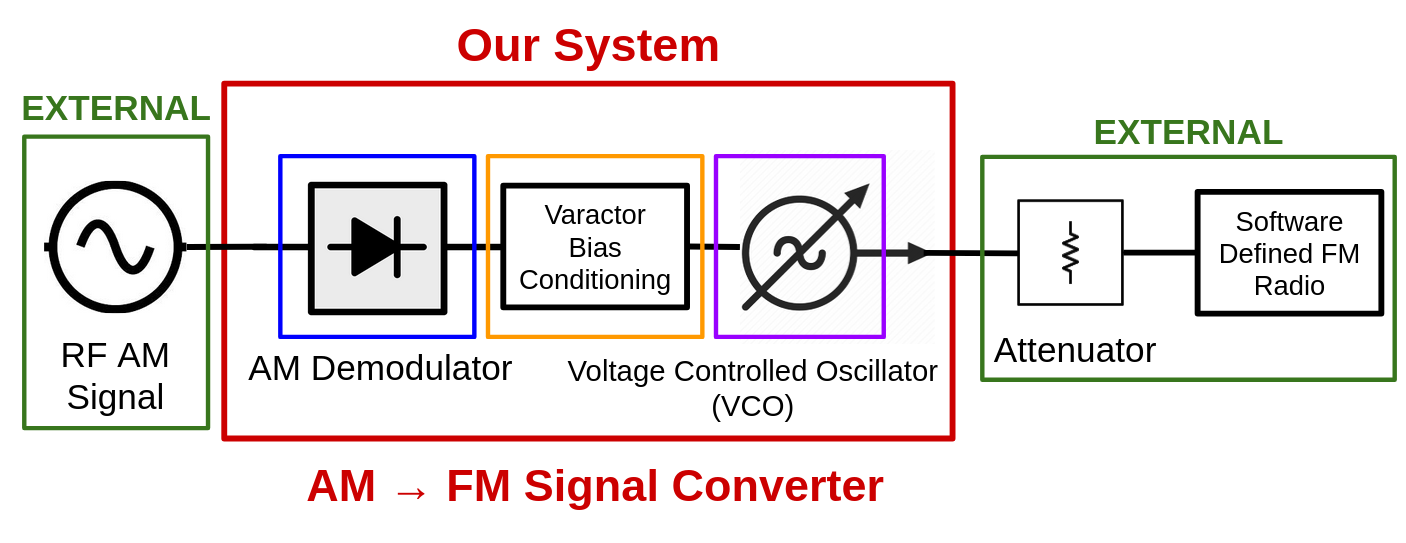}
    \caption{Simple Block Diagram: AM Receiving to FM Broadcasting}
    \label{fig:simple_block_diagram}
\end{figure*}

\begin{raggedright}
At a high level, these blocks are related according to the block diagram in Figure~\ref{fig:simple_block_diagram}. 
\newline
\newline
The AM demodulator receives an AM signal, which is currently the output of an Agilent 33220A, allowing control over a number of parameters, including modulation depth, carrier frequency, signal amplitude, and more. In principle, the Agilent 33220A was utilized so that the limiting factor of our design was not the incoming signal strength, although sub-blocks were still designed for varying signal strengths. The basic task of the AM demodulator is to preserve the envelope amplitude and output a wave at its frequency while ignoring the carrier wave frequency. 
\newline
\newline
Then, this envelope voltage is inputted into the AM-to-FM conversion block---consisting of Varactor Bias Conditioning and VCO---as the eventual control signal for the VCO. Critically, the raw demodulated envelope does not function as the VCO control signal. It must first be attenuated and added to a DC offset such that certain components within the VCO function properly for re-modulation at RF. The block accomplishing this function is the ``Varactor Bias Conditioning". 
\newline
\newline
Finally, the VCO must produce a sine wave, as FM signals are sinusoidal, but at a tunable frequency according to the relative level of our input envelope voltage. In this manner, changes in amplitude are converted to changes in frequency, i.e. going from AM to FM. Technically, this output sinusoidal wave will be an FM signal. However, we must ensure that the signal has the proper characteristics for transmission. This step involves buffering and attenuating the output to increase the output resistance and prevent the output circuitry from impacting the frequency of the oscillator. In this manner, we can preserve listenable AM signals at the input to listenable FM signals at the output. 
\newline
\newline

\end{raggedright}

\begin{raggedright}
For readers interested in signal levels and frequencies at various steps in the signal chain, this diagram describes those and breaks out some blocks within the above diagram into more sub-blocks. The colors align between the diagrams for easier understanding. Only the blocks that are internal to our system are included.  
\end{raggedright}

\begin{figure}[h]
    \centering
    \includegraphics[width=1\linewidth]{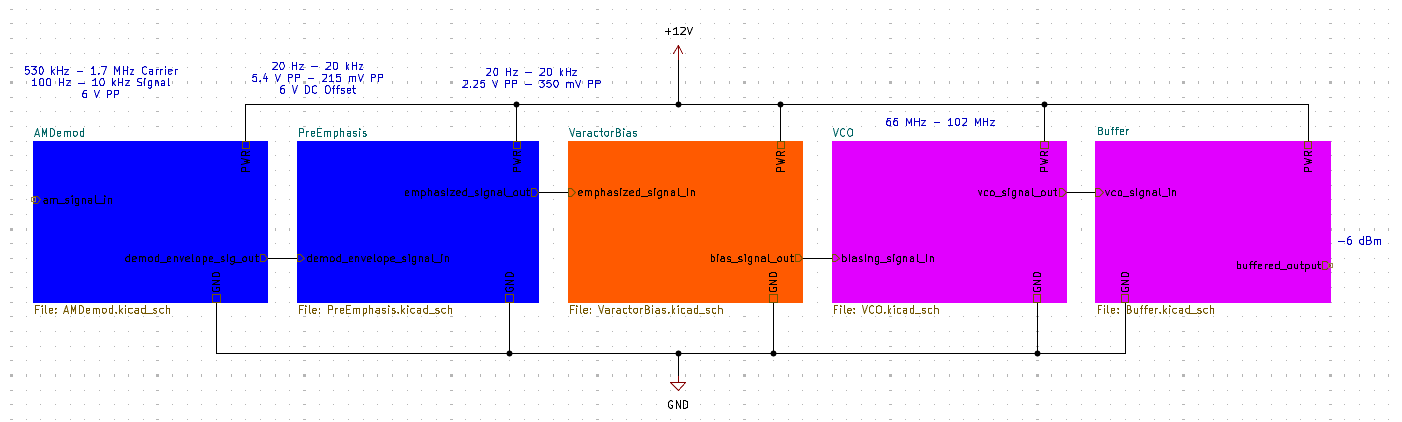}
    \caption{Full Block Diagram: Signal Levels and Frequencies}
    \label{fig:full_block_diagram}
\end{figure}

\begin{raggedright}
Most importantly, the output of our system is at -6 dBm and between 66 MHz and 102 MHz.
\end{raggedright}

\subsection{Design Methodology}
We wanted to build topologies that were robust under a wide range of operating conditions: voltage, frequency, modulation depth, noise, and more. This was best achieved by a combination of mathematical and intuitive thinking coupled with experimentation. For instance, by decoupling the tank circuit of the VCO from the rest of the device, we can easily estimate oscillation frequency using the product of inductor and capacitor values to provide a ballpark for the operating frequency ranges of our circuits. After estimating hand calculations, we simulate all transistor-based circuits, verifying expected bandwidth and gain and ensuring that all transistor-based circuits are effective over the desired operating frequencies. As a final example, circuits are first built on a breadboard to easily estimate performance and experiment quickly with different components, then transferred to a perfboard, limiting breadboard parasitics that plague RF circuits like these.

\subsection{AM Demodulator}
To justify the choice of topology, it is helpful to first discuss other topologies' challenges. The following comes from a discussion given in \cite{lesurf_envelope_detector}. First, we can discuss the simple envelope detector, consisting of simply a diode for rectification and a low-pass RC filter. At its core, the diode preserves the upper half of the incoming signal, and the low-pass RC filter attenuates away the higher carrier frequency, leaving just the envelope signal. There are a number of issues with this working, albeit imperfect, implementation. First, for small signals, i.e. those smaller than the diode's forward voltage drop, the simple envelope detector is useless. The diode never turns on. Also, because we only preserve the top half of the signal, we lose half of our signal. Next, there are trade-offs about pole placement for the RC-filter. Suppose we want to preserve the audible range, which extends to 20 kHz, so we place the pole of our filter at 20 kHz. Then, the AM band begins at 530 kHz, which is about 1.4 decades from the -3 dB point, indicating that if past the -3 dB point, we attenuate at -20 dB per decade, then the attenuation will be by -28.5 dB or a ratio of about 0.04. So, the demodulated signal will still contain this amount of high carrier frequency, a result which can clearly be seen on an oscilloscope. This will show up as a ripple in the demodulated signal, described by $V_{ripple} = \frac{V_{peak}}{f_c \cdot RC}$ where $f_c$ is the carrier frequency. Finally, because the time constant $RC$ of the filtering network is much longer than the carrier wave period, the voltage across the capacitor is supposed to track the input voltage. However, it does not work perfectly. So, we need diodes with fast recovery times to support envelope detectors. At this point, it should be clear that improvements must be made to the envelope detector before it is integrated into our project. Below the an example output of a marginally improved envelope detector---taken from the circuit given in \cite{analog_envelope_detector}. This trial was done with a 600 mV peak-to-peak sine wave carrier at 500 kHz with a 1 kHz modulation at 80\% depth. Already, the resulting ripple on the output is visible. This is not the level of performance we sought to achieve.

\begin{figure}[h!]
    \centering
    \includegraphics[width=0.85\linewidth]{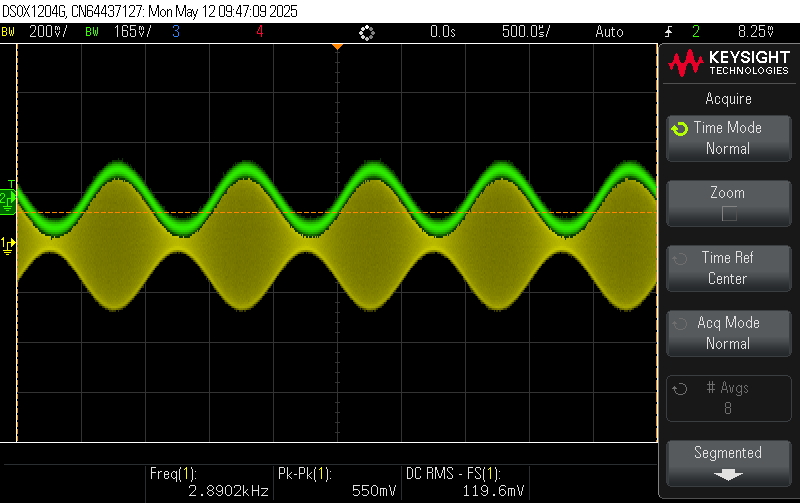}
    \caption{Improved Envelope Detector}
    \label{fig:barely_improved_enevelope_detector}
\end{figure}

\begin{raggedright}
Next, we discuss synchronous modulator topologies such as product detectors. These work by mixing, i.e. creating the sum and difference of frequencies between a local oscillator and the incoming carrier \cite{nickolas_basics_of_mixers}. Specifically, consider that multiplication in the time domain by a phase, i.e. the output of the local oscillator, is equivalent in the frequency domain to convolution. Sinusoids in the frequency domain become Dirac-Delta functions, so multiplication produces the sum and difference of the frequencies. However, product detectors also require that the phase difference between these two signals is near zero \cite{electronics_notes_am_demodulation}. In turn, this generally necessitates the building of a phase-locked loop (PLL) to synchronize these two phases. It is very difficult to build topologies in a short time frame with little prior experience, so although they have better performance than our eventual topology, there was also a chance of the system not working altogether. We decided not to take this risk.
\newline
\newline
Balancing risk and performance, we chose to build a much improved version of an envelope detector topology. For a full characterization of the results, please see the Results section.
\end{raggedright}

\subsection{Voltage Controlled Oscillator (VCO)}
\label{subsec:system_vco}

The VCO has four primary requirements. First, it must produce a sinusoidal output to avoid spurious frequency generation. Second, it must produce a signal around 100 MHz to be within the FM broadcast band. Third, it must be able to be tuned across at least 5 MHz so that it can be placed at a frequency that does not overlap with existing FM radio stations. Finally, fourth, it must be able to modulate the output frequency for an input signal up to 10 kHz so that the majority of the audible spectrum---stretching from 0 Hz to 20 kHz---can be encoded. The output frequency and sinusoidal requirements eliminate many oscillator configurations, such as those based on the 555 Timer, which has a rise and fall time of around 100 nanoseconds (ns), limiting its maximum frequency to 5 MHz \cite{555datasheet}. For this reason, as well as the requirements for 5 MHz tuning range, we chose the Colpitts Oscillator, which as an LC oscillator exhibits all of these features \cite[Table 7.2]{Horowitz:1981307}. Similar LC oscillators could likely achieve better performance. However we chose to start with a simple oscillator due to the unpredictability of working with high frequencies and the short development time frame. Additionally, the choice of a Colpitts Oscillator over a Hartley Oscillator was made due to the greater availability of capacitors compared to inductors, and the lower parasitics of a typical ceramic capacitor.

\section{Circuit Design}

Whereas before we sought to define design goals or justify entire topologies, this section will focus on specific circuit modules---from the AM demodulator to the output buffer---and why particular sub-topologies were utilized and how specific components were sized.

\subsection{AM Demodulator}

This section focuses on a sub-block and component-level design choices within the AM Demodulator. It follows from an improved envelope detector first developed in the given source \cite{schenck2002improved}. The entire circuit uses a virtual ground at 6 V so the op-amps can swing from 0 V to 12 V using a single power supply.

\subsubsection{Full Wave Precision Rectifier}

In envelope detectors, we must first employ a rectifier. The envelope detector is essentially a voltage hold circuit, where the discharge time of the capacitor across the resistor is much longer than the carrier frequency, i.e. $f_c >> \frac{1}{2\pi \cdot RC}$. So, the capacitor seems to hold the peak voltage of the incoming wave, corresponding to the envelope signal. However, consider the case where we have no rectification, meaning the capacitor sees equal times of positive and negative voltages. During the positive cycles, the capacitor would have a positive voltage at its upper terminal with regard to its lower terminal, and so discharge across the resistor. During the negative cycles, it would have a negative voltage at its upper terminal with regard to its lower terminal. The overall output would be equally positive and negative across the resistor, averaging to zero. This is obviously a problem. So, rectification is necessary to isolate either only the top or bottom half of the incoming waveform: $v_+(t)$ or $v_-(t)$. However, with half-wave rectification, the corresponding half of the waveform is disregarded entirely, which is why we build a full-wave rectifier.
\newline
\newline
Next, precision rectifiers employ operational-amplifier (op-amp) feedback to improve traditional rectifiers. Under the Ideal Op-Amp Assumption, the positive and negative input terminals are at the same voltage, and no current flows into or out of these terminals. So, if we have a configuration like that shown below, focusing on only one of the diodes, the output is a one-to-one with the input ($R_{in} = R_{feedback}$), albeit only the positive half of the wave. The 0.1 uF capacitor presents negligible impedance to signals even of our lowest chosen frequency ($\frac{1}{2\pi fC} = \frac{1}{2\pi (530 \cdot 10^3)(0.1 \cdot 10^{-9})} \approx 3 \Omega$). The exact component value is not as important as that it presents a large impedance to direct current (DC) signals. Most importantly, in precision rectifiers, the signal does not have to overcome the forward voltage drop of the diode---op-amp feedback takes care of that by acting as an almost infinite differential amplifier between its input terminals. 
\newline
\newline
For high frequencies, diodes have recovery times due to regions of charge forming within them. This can cause distortion as the signal crosses over the negative threshold. Hence, the 1N4148 were chosen because their recovery times are $\approx 4$ ns and $4 \text{ns} << \frac{1}{f_{max}} = \frac{1}{1.7 \text{MHz}} \approx 588 \text{ns}$ \cite{1N4148datasheet}. Additionally, the LM318 op-amp was chosen, with a gain-bandwidth product (GBP) of 15 MHz, as it was the closest available part suited to RF, with open-loop gain of 15 at 1 MHz \cite{LM318datasheet}. Rigorously, Black's Formula tells us that: $G_{cl} = \frac{G_{ol}}{1 + f\cdot{}G_{ol}}$. Hence, when $G_{ol}$ is assumed to be very large, the closed-loop gain is determined by $f$, the feedback term, which for this case is 1. So, $G_{cl} = \frac{15}{16} \approx f$ at 1 MHz, even if it is not perfect. There are very specific op-amps with GBPs in the hundreds, but they are comparatively expensive. Finally, even driving rail-to-rail at 6 V, we are not concerned with the slew rate of the LM318, which is typically 70 V per microsecond \cite{LM318datasheet} because $f_{max} = \frac{SL}{2\pi \cdot V_p} = \frac{70 \cdot 10^6}{6 \cdot 2 \pi} \approx 2.2 \text{MHz} > 1.7 \text{MHz}$. So, any op-amp with a higher slew rate is not needed. Anyways, we would not be driving the LM318 rail-to-rail because it is not a rail-to-rail op-amp \cite{LM318datasheet}.
\newline
\newline
Below is the schematic of our full-wave precision rectifier:
\begin{figure}[h]
    \centering
    \includegraphics[width=0.75\linewidth]{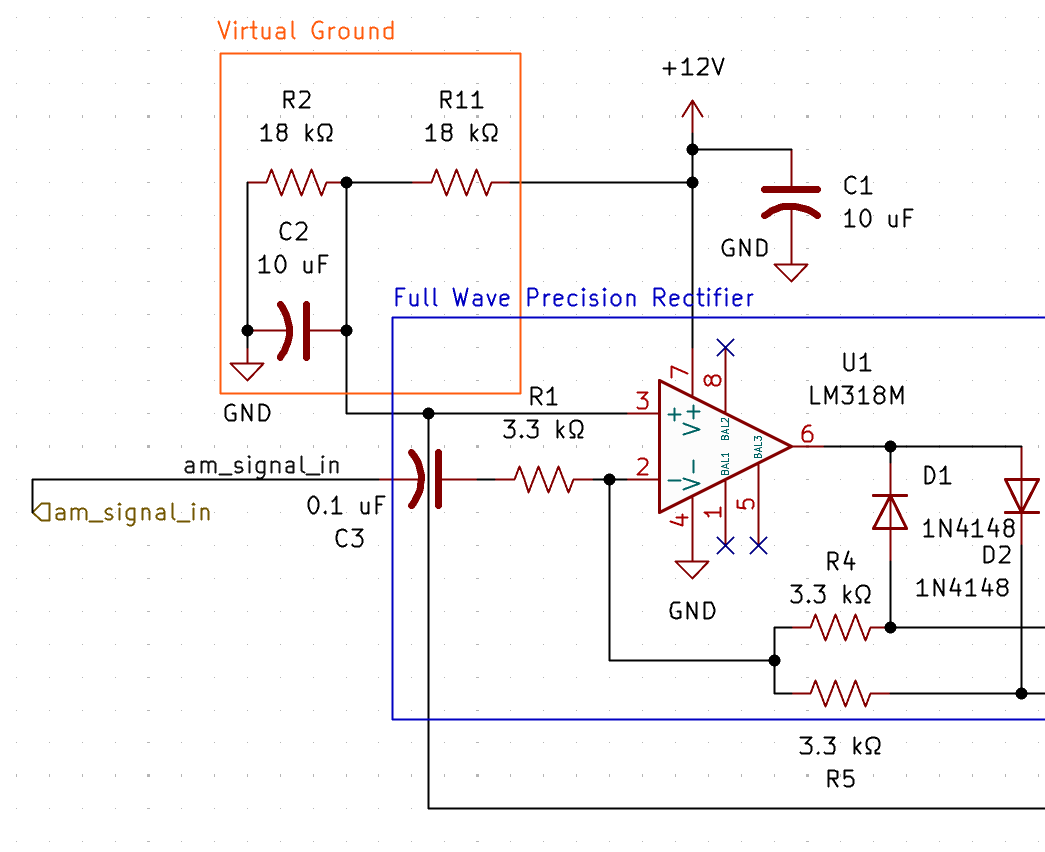}
    \caption{Full Wave Precision Rectification Topology}
    \label{fig:precision_rectifier}
\end{figure}

\subsubsection{Wave Re-Combination}

After precision rectification, we are left with the positive and negative halves of the waves. We can invert the positive half of the wave and sum the positive and negative halves. The positive half of the wave---coming from the diode marked D2 on the schematic---is fed into a non-inverting amplifier with gain $G = 1 + \frac{R_8}{R_7} = 2$. Then, the output voltage, which is twice the positive half of the wave, is fed into another non-inverting op-amp topology where the non-inverting terminal is the negative half of the wave, again scaled by gain $G = 1 + \frac{R_6}{R_9} = 2$. Hence, the positive half of the wave is flipped and added to the negative half, both halves having had a gain of 2 applied. In the schematic on the next page, the lower op-amp handles the positive half of the wave, and the upper op-amp handles the negative half of the wave. Since this stage still deals with the carrier frequency, the op-amps must have a sizable gain-bandwidth product, so LM318s were again used. Overall, the choice of gain is largely unimportant, but signal levels should not be allowed to be excessively low. 

\newpage

\begin{figure}[h!]
    \centering
    \includegraphics[width=0.25\linewidth]{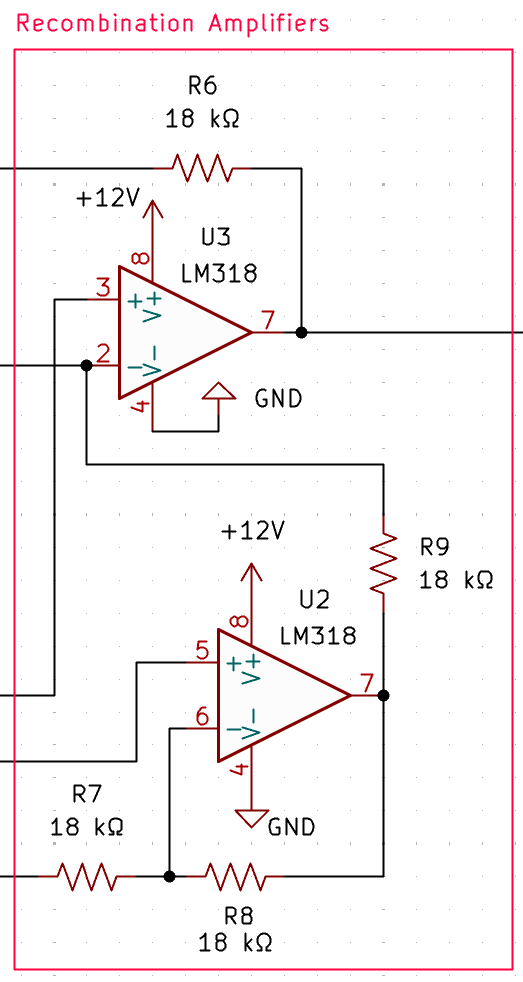}
    \caption{Recombination Amplifier Topology}
    \label{fig:recombination}
\end{figure}

\subsubsection{Output Filter}

The output filter is simply an RC filter with a time constant of $\tau = (10^4)(22 \cdot 10^{-9})$, making the cutoff frequency $f_{cutoff} = \frac{1}{2\pi \cdot RC} = \frac{1}{2\pi \cdot \tau} \approx 723 \text{Hz}$. This may appear to be over-filtering the audible band, but empirical experimentation shows that over-filtering reduces some of the nonlinearities present with the LM318s due to the Ideal Op-Amp Assumption being invalid for a $G_{ol} = 15$ at 1 MHz.

\begin{figure}[h!]
    \centering
    \includegraphics[width=0.65\linewidth]{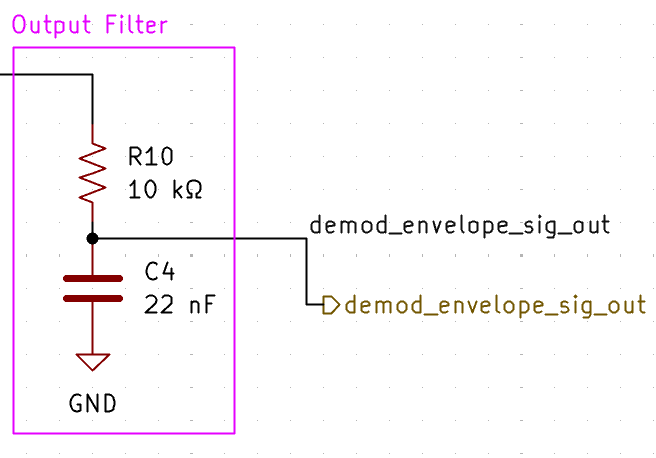}
    \caption{Output Filter Topology}
    \label{fig:output_filter}
\end{figure}

\subsubsection{Schematic}

The schematic below shows the full AM demodulator, detailing how the sub-blocks are connected together. To reiterate, first the incoming signal is precision rectified, then recombined with a series of non-inverting amplifiers, and finally filtered with a first-order RC low-pass filter. The result is a sine wave with the same frequency as the input modulating wave's frequency.

\begin{figure}[h!]
    \centering
    \includegraphics[width=\linewidth]{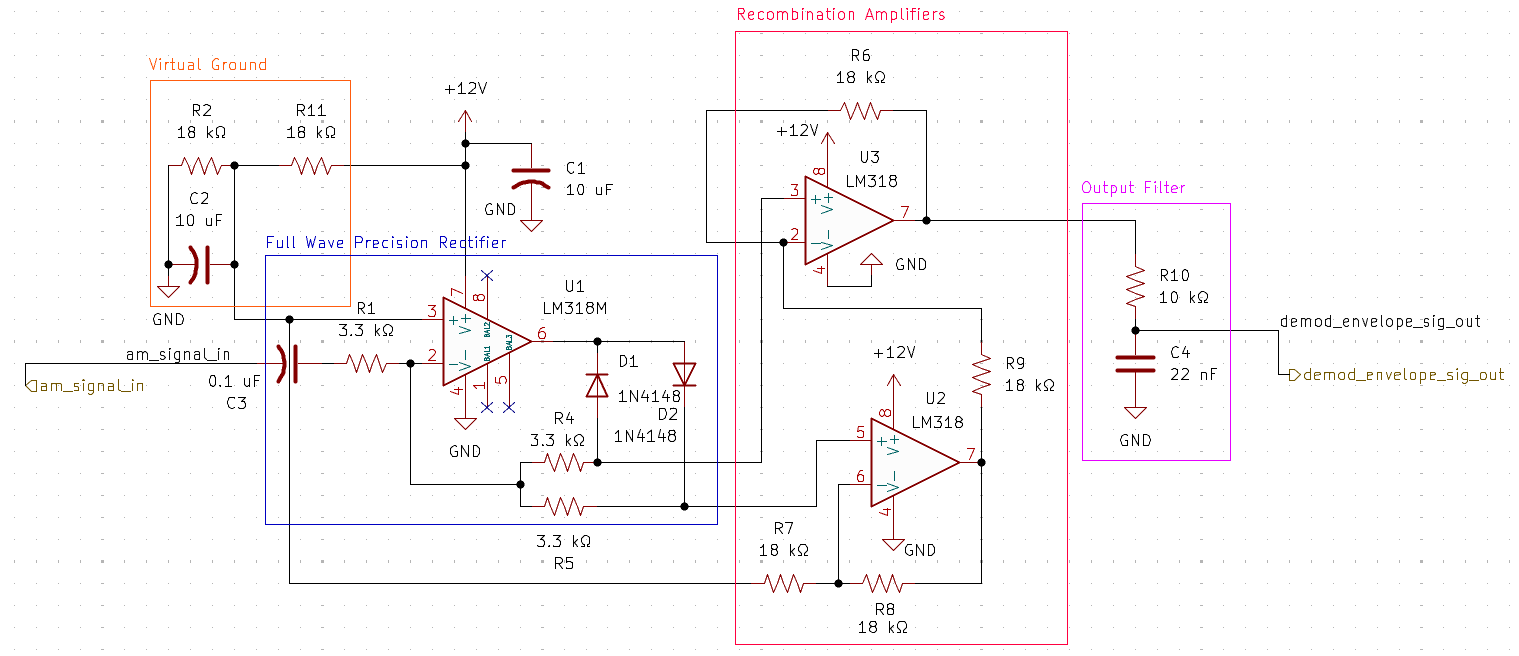}
    \caption{Full AM Demodulation Topology}
    \label{fig:am_demodulator}
\end{figure}

\subsubsection{Simulations}

Before constructing the AM demodulator, we created LTspice simulations which allowed us to visualize signals at multiple nodes within the circuit and accurately tune values. For this simulation, because LTspice does not possess a native model for the LM318 op-amp, we used a universal op-amp, making sure to set the gain bandwidth product to 15 MHz---the most important performance specification for the LM318. Below is a schematic showing the LTspice simulation set-up. The carrier is a 0 V DC offset, 1 $V_{pp}$, and 1 MHz, and the signal is 1 V DC offset, 1 $V_{pp}$, and 1 kHz. As can be seen in the simulation output, this corresponds to 100\% modulation depth.

\begin{figure}[h!]
    \centering
    \includegraphics[width=0.7\linewidth]{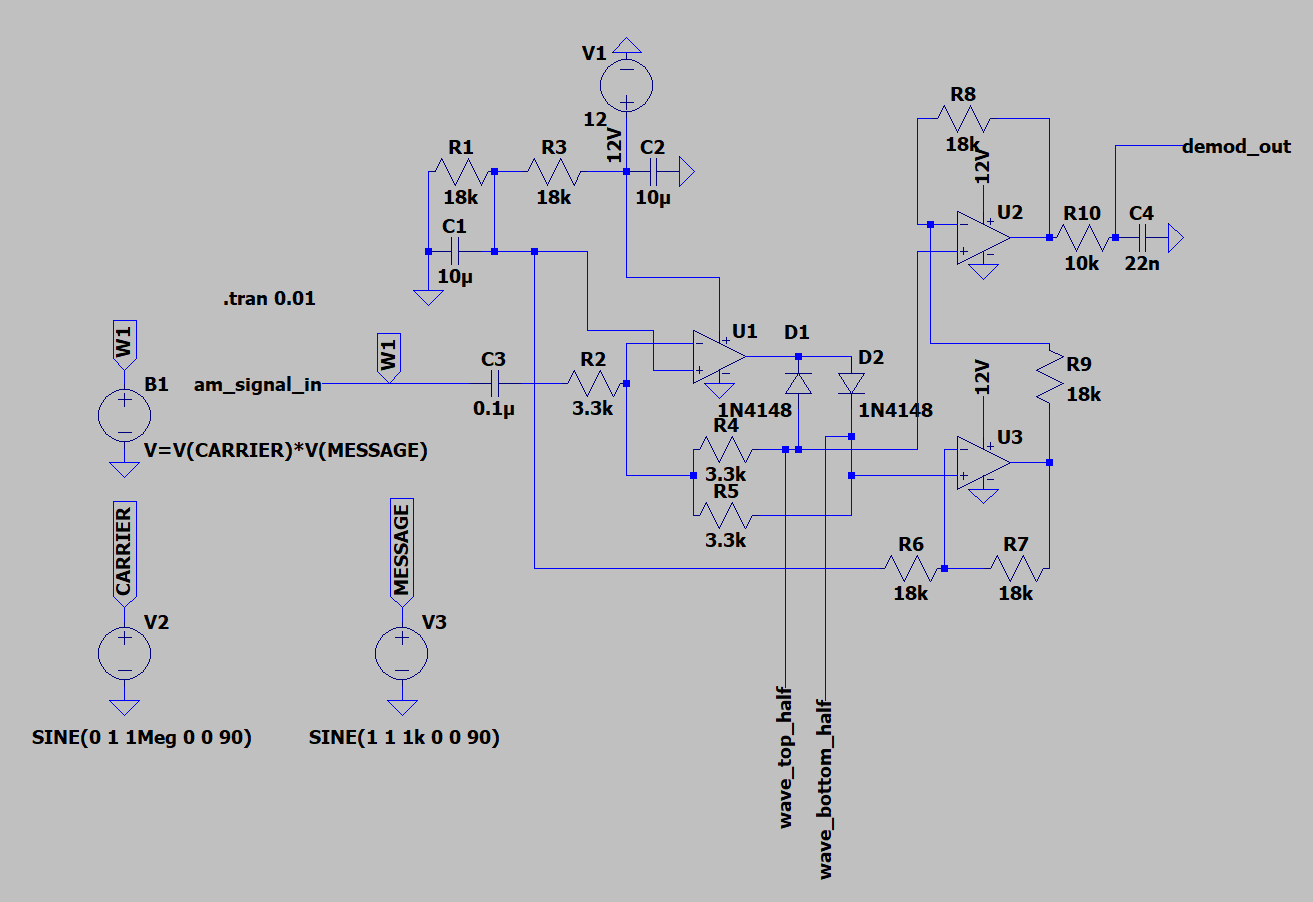}
    \caption{AM Demodulator Simulation Setup}
    \label{fig:am_demod_sim_setup}
\end{figure}

\begin{raggedright}
For the simulation, we probe the input, output, and upper and lower halves of the wave. We want an output that is a 1 kHz sine wave. The simulation output below shows that this is accurate. Because the labels are difficult to read: green is the input, red is the top half of the wave, blue is the bottom half of the wave, and turquoise is the output.
\end{raggedright}

\begin{figure}[h!]
    \centering
    \includegraphics[width=0.75\linewidth]{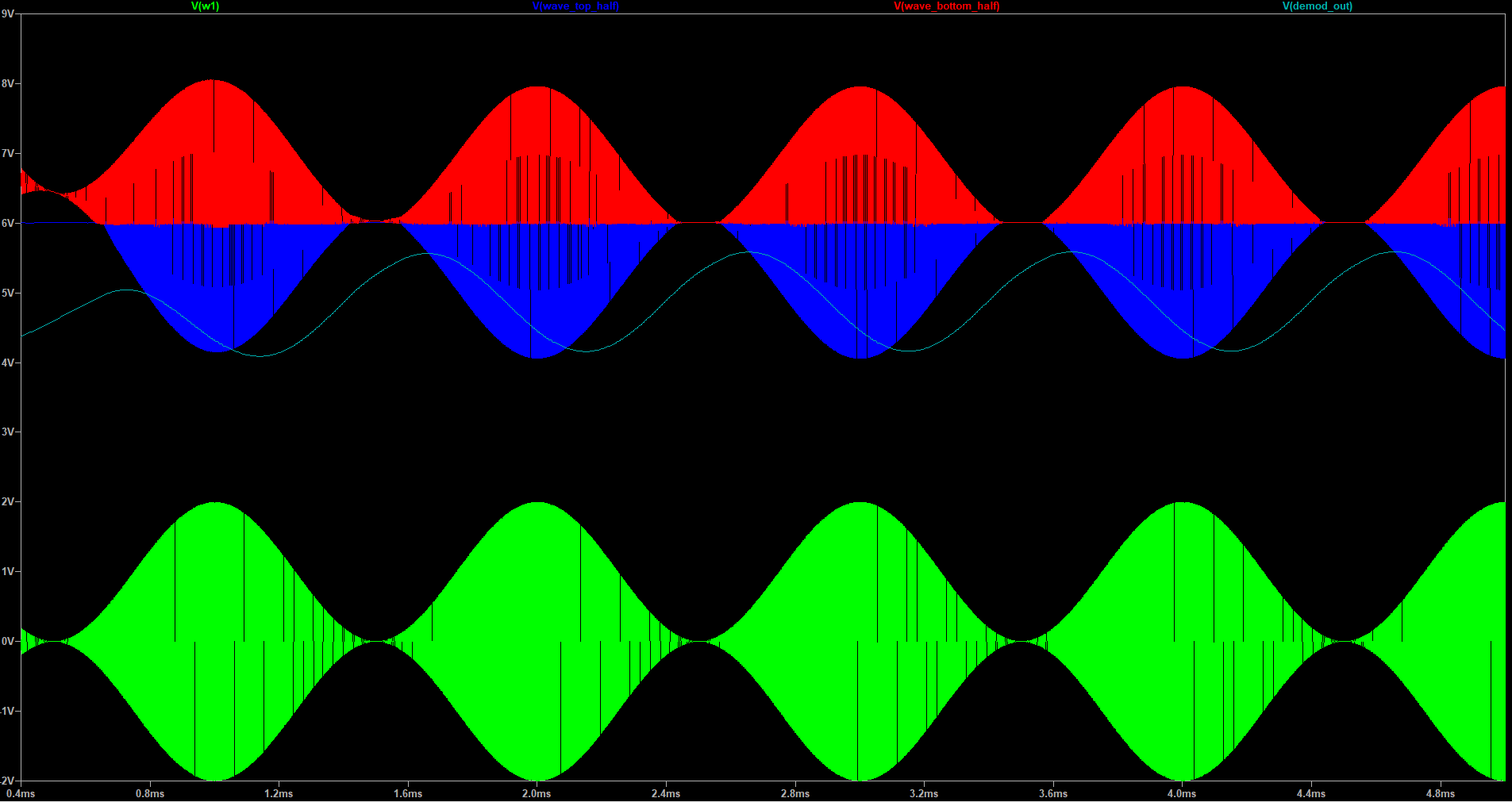}
    \caption{AM Demodulation Transient Simulation}
    \label{fig:am_demod_transient_sim}
\end{figure}

\begin{raggedright}
Additionally, because we seek a uniform frequency response from the AM Demodulator across the AM band, from 530 kHz to 1.7 MHz for the carrier, and similarly uniform from 100 Hz to 10 kHz for the message, we can perform an AC sweep across those values, observing the output gain and phase. In the LTspice simulation itself, nothing changes except that the carrier or message signal becomes the AC sweep input, configured to start and end at each end of the range, going by decade with 10 points per decade. Of course, we expect high frequencies to be attenuated downward due to the pole-placement of the output filter, but they should not be attenuated to unreasonable levels. Additional simulations confirmed that the amplitudes at not overly attenuated at either side of the frequency range, both for the carrier frequencies and modulation frequencies. Additional diagrams are not included due to space constraints. For instance, 5 kHz input signals produce 5 kHz output signals on the order of 50 m$V_{pp}$. With the following pre-emphasis circuit providing gain to high frequencies, this is not concerning.

\end{raggedright}

\subsection{Pre-Emphasis Circuit}
In this section, we discuss the pre-emphasis circuit, which attempts to correct the attenuation of high frequencies from the AM demodulation topology. 
\subsubsection{Schematic}

On the next page, we have the simple schematic of the pre-emphasis circuit.

\begin{figure}[h]
    \centering
    \includegraphics[width=\linewidth]{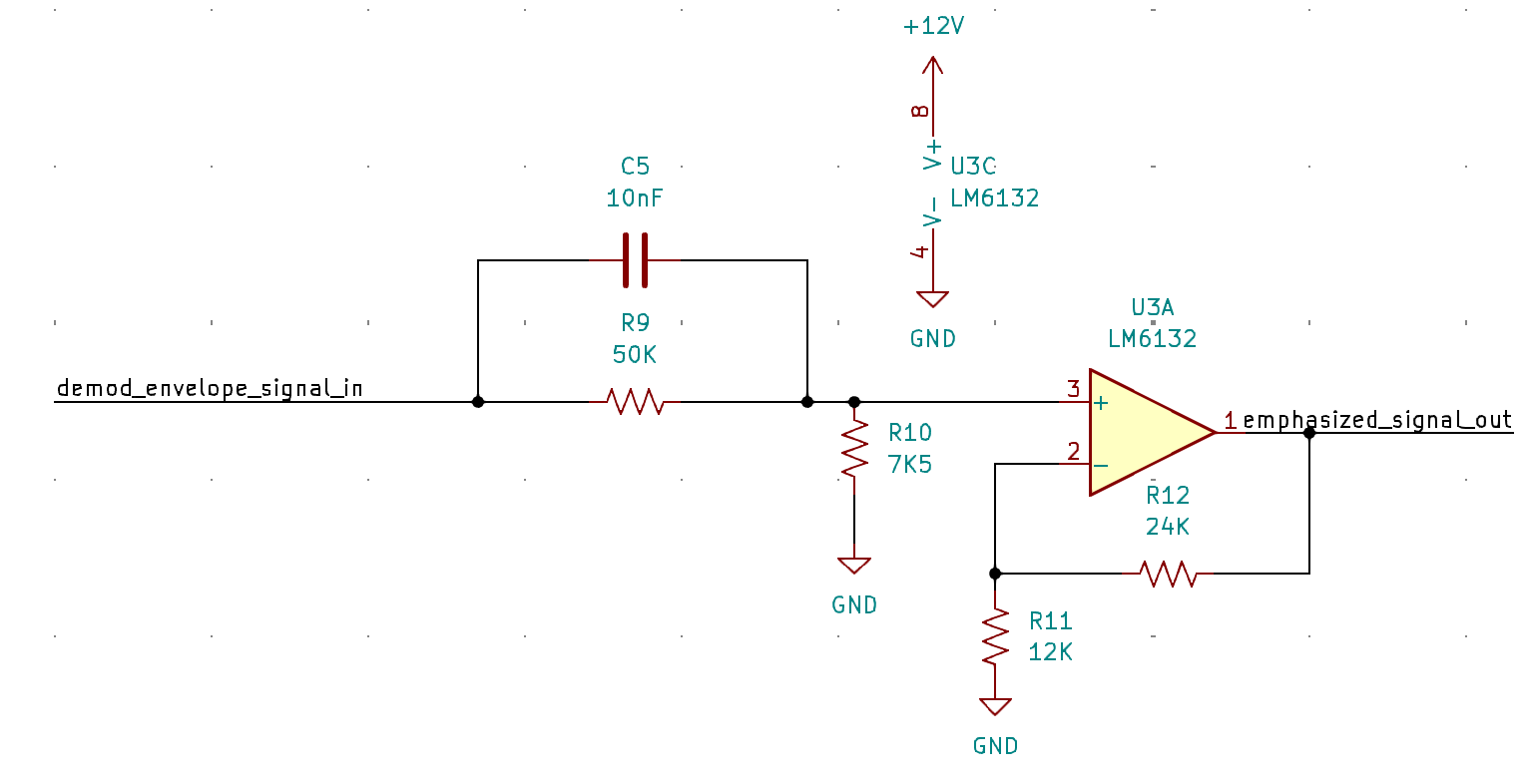}
    \caption{Pre-Emphasis Topology}
    \label{fig:pre_emphasis}
\end{figure}

\subsubsection{Theoretical Analysis}

From the theoretical analysis of the AM demodulation topology, we see that the output filter has a time constant of $\tau = 220 \mu s$, which corresponds to a frequency of 723 Hz. However, this means that frequencies in the audible band are also being attenuated, meaning that if the raw outputs from the AM demodulator were used to play music, it would have overly prominent low frequencies. In other words, the music would be bass-boosted.
\newline
\newline
To solve this problem, we develop a pre-emphasis circuit that resolves the attenuation in the audible frequencies while still attenuating at high frequencies. At frequencies below the lower crossover frequency of $\tau_1 = (10  \text{ nF})(50 \text{ k}\Omega) = 500 \mu s$ and $f_1 = 318 \text{ Hz}$, the signal is attenuated, while at frequencies above the upper crossover frequency of $\tau_2 = (10  \text{ nF})(7.5 \text{ k}\Omega) = 75 \mu s$ and $f_2  = 2.1 \text{ kHz}$. Between $f_1$ and $f_2$, the frequency response increases at a constant 20 dB/decade, which counteracts the attenuation from the AM demodulation. In essence, this pre-emphasis circuit, when combined with the AM demodulator, amplifies signals from 318 Hz to 723 Hz in the middle of the audible band for music. Then, frequencies from 723 Hz to 2.1 kHz are no longer attenuated, and lower frequencies are also no longer boosted. In combination with an op-amp connected in a non-inverting topology with a gain of 3, we find that the pre-emphasis circuit cleans up the signal from our AM demodulation as desired.

\subsubsection{Simulation}

\begin{figure}[H]
    \centering
    \includegraphics[width=0.9\linewidth]{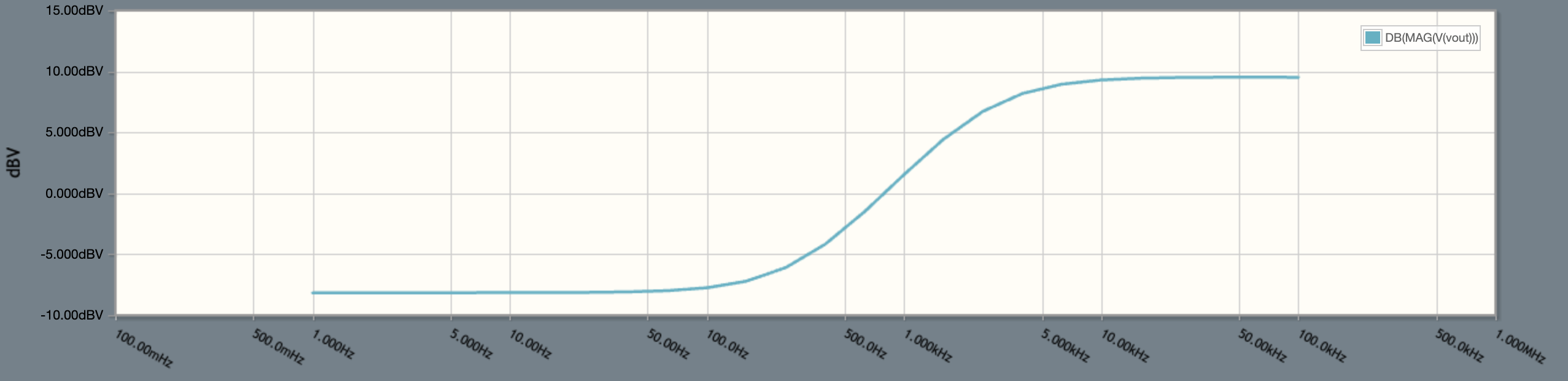}
    \caption{The simulation of the pre-emphasis circuit, demonstrating expected turning points.}
    \label{fig:preemph-sim}
\end{figure}

We simulate the pre-emphasis circuit using SPICE, finding that the turning points in the simulated curve are as expected, with -3dB points around 300 Hz and 2 kHz. As such, we validate the hand calculations using simulation.

\subsection{Voltage Controlled Oscillator}
In this section, we focus on the function of the voltage controlled oscillator. 

\subsubsection{Schematic}

\begin{figure}[H]
    \centering
    \includegraphics[width=\linewidth]{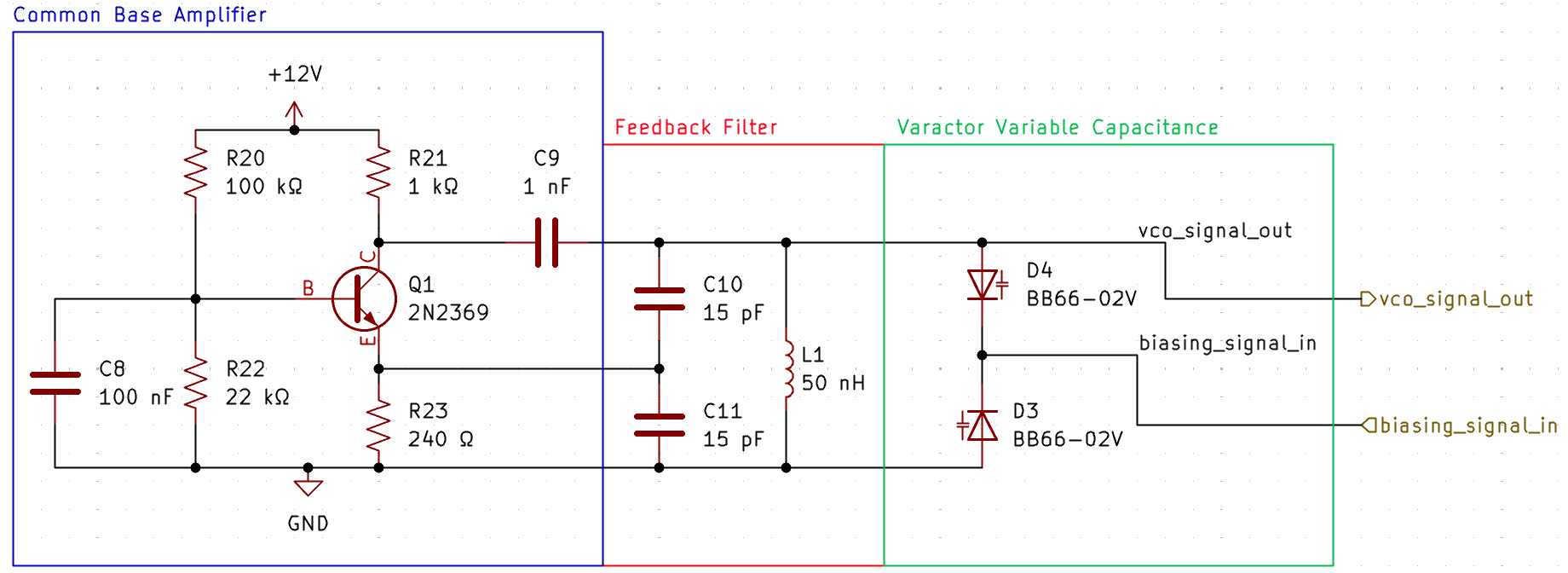}
    \caption{Colpitts Oscillator Topology}
    \label{fig:colpitts_oscillator}
\end{figure}

\subsubsection{Theoretical Analysis}



\begin{raggedright}
\label{subsubsec:colpitts_analysis}
The Colpitts Oscillator operates under the principle of frequency-dependent positive feedback. An amplifier is connected in a loop with a bandpass filter, amplifying only frequencies that are within the filter's passband. While signals outside of this passband decay, the passed signals are amplified until the signal reaches an amplitude large enough that the swing of the operating point effectively reduces the gain of the amplifier to 1. If the loop gain is exactly 1 and the phase shift around the loop is a multiple of $360^\circ$ for a given frequency, the Barkhausen Criteria are satisfied and oscillation may occur at this frequency \cite[pp. 449]{Horowitz:1981307}.
\newline
\newline
The design of this oscillator was driven by the challenges presented by the high operating frequency. Even the 2N2369 transistor, a transistor ``designed for high-speed switching applications" has an input capacitance around 4 pF \cite{2369datasheet}. This value is quite large compared to the $\approx 15 $ pF equivalent capacitance present in the bandpass filter, and would significantly hamper the ability of the oscillator to operate at high frequencies. To mitigate this, a common base amplifier was used, which minimizes the effects of the transistor's internal capacitance. While there is still some capacitance from input to output, the amplifier is non-inverting, so the amplifier does not experience the Miller effect which would increase the effective capacitance.
\newline
\newline
The operation frequency is set by the resonant frequency of the feedback filter. The resonant frequency for an LC tank circuit is $1/(2\pi\sqrt{LC})$. The varactor diodes, which act as electrically-tunable capacitors, have a minimum capacitance of 12 pF, which yields 6 pF in their series combination. With an inductance of 50 nH and combined and total combined capacitance $C = 13.5pF$ (both 15 pF capacitors and varactor diodes) we find $f_0 = 193.7 \, \mathrm{MHz}$. In reality, this frequency of operation is much lower, around $100 \, \mathrm{MHz}$, likely due to parasitic capacitance in the physical circuit construction and parasitics in the components themselves.
\newline
\newline
To make this oscillator voltage controlled, we can modulate the capacitance, and therefore the resonant frequency, of the feedback filter using the varactor diodes. The diodes have a capacitance that is dependent on the reverse bias voltage, so by modulating the reverse bias voltage, we modulate the frequency of the oscillator. A more detailed analysis of the biasing of these diodes can be found in section \ref{subsec:varactor_bias}.
\end{raggedright}

\subsubsection{Simulation}
The VCO was simulated in LTspice to verify its function before construction. We also used simulation to verify our intuition when tweaking the circuit between the many design iterations that were necessary to achieve the desired operation frequency. Our initial simulations showed that the device should oscillate at 150 MHz with an amplitude of $300 \, mV_{pp}$ with similar capacitors but an inductance of 200 nH. However, the constructed device did not oscillate, likely due to losses from parasitics that the simulation did not account for. 

\subsubsection{Design Iteration}
The VCO required many iterations to achieve a sufficiently high operating frequency. Our first iteration was on a breadboard, oscillating around 20 MHz. The first modification we attempted was switching the transistors from 2N3904 to the higher transition frequency transistors of the 2N2369 series, which allowed us to decrease the inductor sizing and increase oscillation frequency to 50 MHz. For our next iteration, we had to further increase the gain of the common base amplifier at high frequencies. 
\newline
\newline
To do this, we decreased the resistor values on the common base amplifier to 1 k$\Omega$ and 240 $\Omega$, keeping the ratio constant to keep the same voltage biasing conditions, and thus the same overall gain. However, with lower resistances (and thus higher load currents), we found that the frequency response had a turning point at a higher frequency, resulting in a better frequency response.

\subsection{Varactor Diode Biasing Circuit}
\label{subsec:varactor_bias}
In order to modulate a signal onto the carrier generated by the VCO, it must be appropriately scaled to achieve the desired bandwidth, and it must include a DC component to set the carrier frequency. The DC component is the bias voltage of the varactor diodes, which sets the frequency of the oscillator when there is no input frequency. On top of this, the small AC signal creates slight deviations in the bias voltage, and therefore the frequency, to perform the signal modulation. This is achieved with a non-inverting adder that takes input from 10-turn potentiometers, allowing the user to control both the DC bias and AC signal level.

\subsubsection{Schematic}
\begin{figure}[H]
    \centering
    \includegraphics[width=\linewidth]{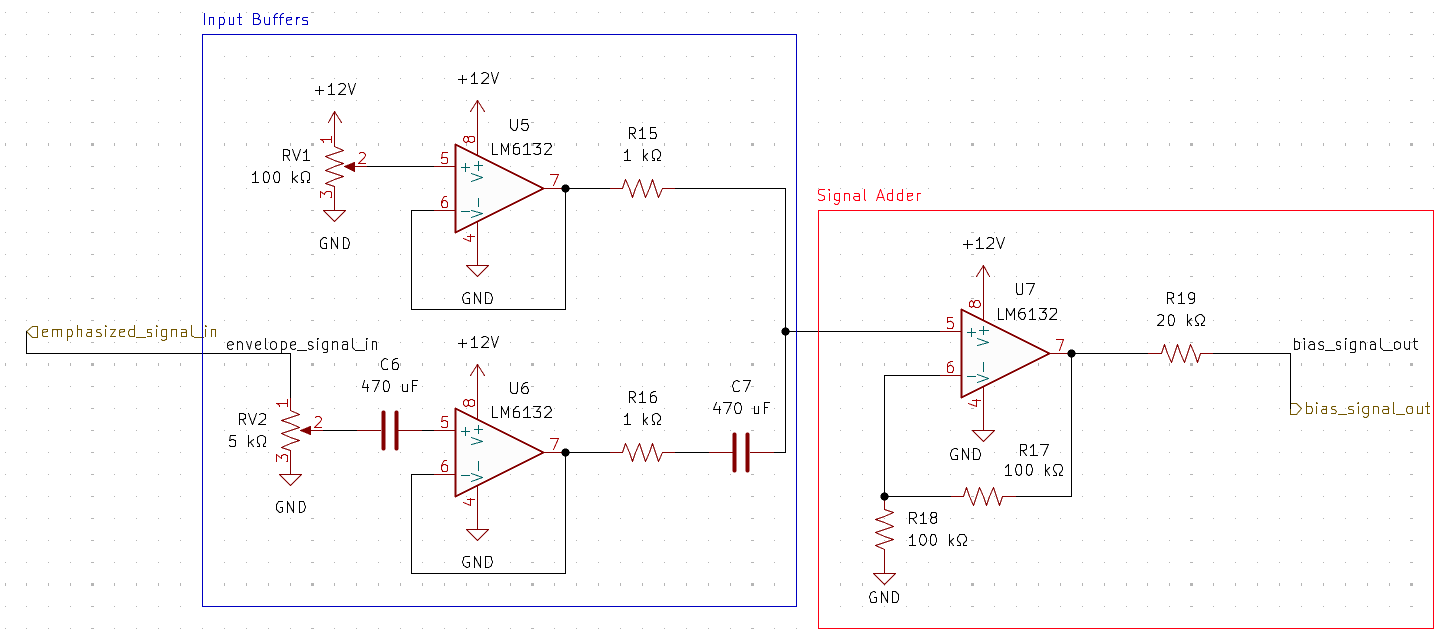}
    \caption{Schematic of the non-inverting adder biasing circuit.}
    \label{fig:varactor_bias}
\end{figure}

\subsubsection{Theoretical Analysis}
A potentiometer is used to generate a DC signal between 0 and 12V. Similarly, a potentiometer attenuates the AC signal, which is then AC-coupled so that the DC level of this signal does not affect the bias point. By buffering both of these signals, they are made more robust to external factors, eliminating another source of sensitivity for the system. To analyze the adder, we will make the ideal op-amp assumption, which gives a gain of 2 for the non-inverting amplifier. Also, the capacitors can be treated as shorts at AC, because they feature a low impedance of $Z = 1/j\omega C=-3j \, \Omega $ at 100 Hz, a frequency at the lower end of the audible range.  Now, it can be seen that the input voltage to the non-inverting amplifier is equal to the DC signal plus half of the AC signal, which is then amplified by a gain of 2 to produce the final varactor diode bias signal. Importantly, this is connected to the varactor diodes through a $20$ \text{k}$\Omega$ resistor, which ensures that the oscillator will not be disrupted. Unfortunately, the resistor in combination with the capacitance of the varactor diodes forms a low-pass filter that could potentially distort our input signal. With $R = 20 \, \text{k}\Omega$ and $C \approx 100$ pF, we get a cutoff frequency $f_c = 1/(2\pi RC) = 79.6$ kHz, which is much higher than audible frequencies. Therefore, this effect can be safely disregarded.

\subsection{Output Buffer}
This section will discuss the output buffer, which is necessary to prevent any loads, especially reactive loads, from changing the transfer function of the feedback filter.

\subsubsection{Schematic}

\begin{figure}[H]
    \centering
    \includegraphics[width=0.75\linewidth]{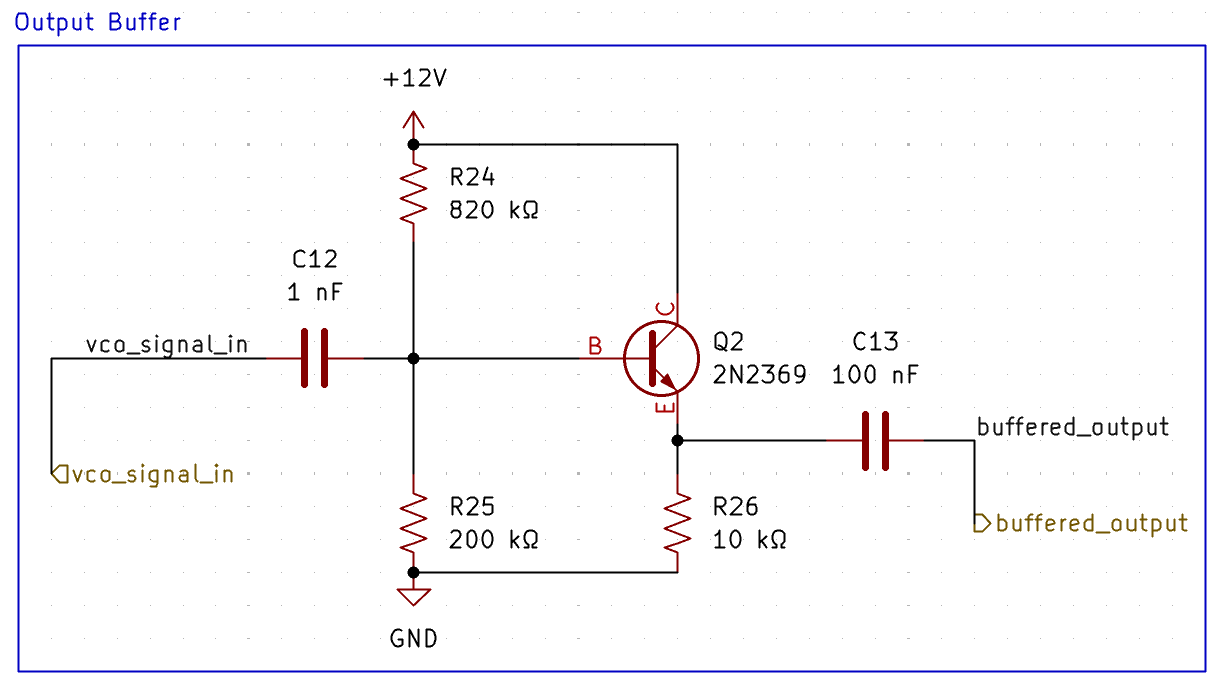}
    \caption{Output Buffer Topology}
    \label{fig:output_buffer}
\end{figure}

\subsubsection{Theoretical Analysis}
Using a small-signal analysis of the transistor, we find that the gain of the emitter follower can be calculated as $a_v = \frac{1}{1 + \frac{R}{R_E}}$, where $R = \frac{R_S + r_\pi}{\beta + 1}$. Given the large value of $R_E$ used, we expect that this gain approaches one, emphasizing the behavior of the emitter follower as a buffer.
\newline
\newline
Using similar small signal analyses, the input resistance of the emitter follower is $r_\pi + (\beta_0 + 1)R_E$ and the output resistance is $R_E || \frac{r_\pi + R_S}{\beta_0 + 1}$, meaning that the input resistance is high (on the order of megaohms) but the output resistance is low (on the order of ohms), making the emitter follower a good buffer.
\subsubsection{Simulations}
\begin{figure}[H]
    \centering
    \includegraphics[width=0.9\linewidth]{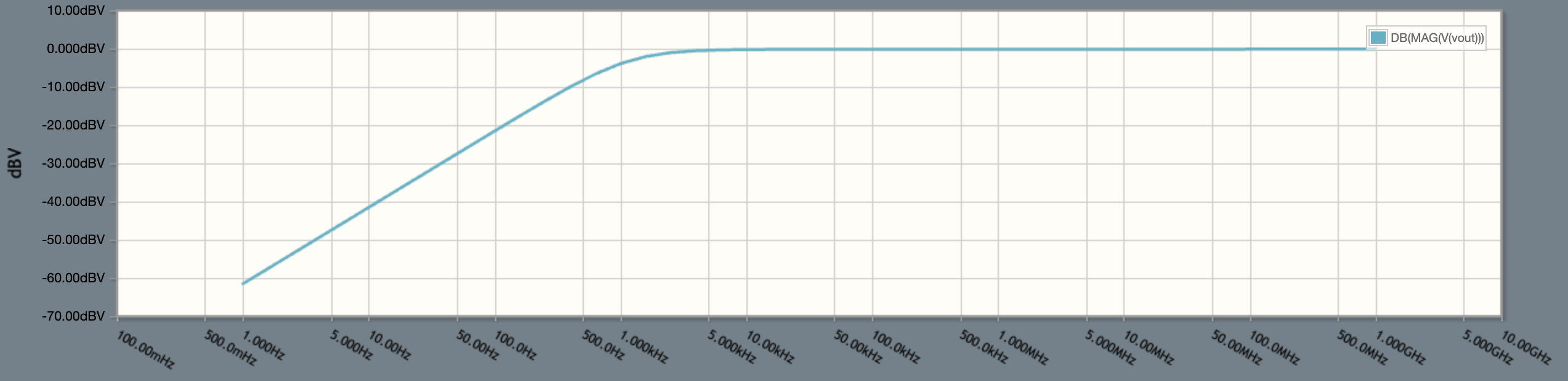}
    \caption{Simulation of emitter follower frequency response, showing gain of 1 in the FM band.}
    \label{fig:EF-sim}
\end{figure}
We simulate the output emitter follower to find the gain, and we find that the gain of the emitter follower is one in the entire AM band around 100 MHz. This means that we can treat the emitter follower effectively as a buffer in this situation, satisfying our requirements.

\subsection{Real-World Circuit Design Considerations}
The above circuits all evaluate our design from a theoretical and simulation-based perspective, which is useful for initial hand calculations, but not exactly matched to the quality of the final system. For example, at the capacitance values we are using of 15 picofarads, the breadboard inter-rail parasitic capacitance of three to four picofarads is no longer negligible. Hence, breadboard was treated as a testing ground for us, where initial circuits could be prototyped to evaluate whether fundamental behaviors are seen, such as oscillation of the circuit, or envelope detection, before being ported over to perfboard.

\section{Results}

This section will discuss our final system's empirical results. It will include measurements for important characteristics of the system---determined based on our project goals. Some of these goals are quantitative, and others are qualitative due to the difficulty of evaluating them quantitatively. For the latter, a good example might be visually comparing the output of the AM demodulator to a sine wave, but evaluating that quantitatively is much more difficult.

\subsection{Submodule Characterization}

In this section, we discuss the empirical characterization of the different submodules. These modules should perform according to their design choices. So, for instance, the VCO's oscillation frequency should be within the FM band---from 88 MHz to 108 MHz---and the AM demodulator's output with a sine wave input should similarly be a sine wave output.

\subsubsection{AM Demodulator}

As mentioned previously, the AM demodulator was specifically designed to be resilient to a host of different operating conditions: modulation depth, carrier frequency, envelope frequency, and more. Unfortunately, the DSOX1204G cannot provide a frequency response analysis (FRA) where the input signal is AM. So, we must alter the demodulator's operating parameters and observe the resulting change in circuit outputs to evaluate whether the demodulator performs over the expected range. In seeking to characterize our demodulator, we experimented with carrier frequencies across the AM band---from 600 kHz to 1.7 MHz---and modulation frequencies across the audible spectrum---from 500 Hz to 2.1 kHz. Because an automated characterization using the oscilloscope is limited, images visualizing a few outputs must suffice.
\newline
\newline
For the first set, we alter the modulation frequency from 500 Hz to 2.1 kHz while holding other parameters constant, observing in both cases that the output in green has a sinusoidal shape with the same frequency as a reference signal in blue.

\begin{figure}[H]
    \centering
    \includegraphics[width=0.85\linewidth]{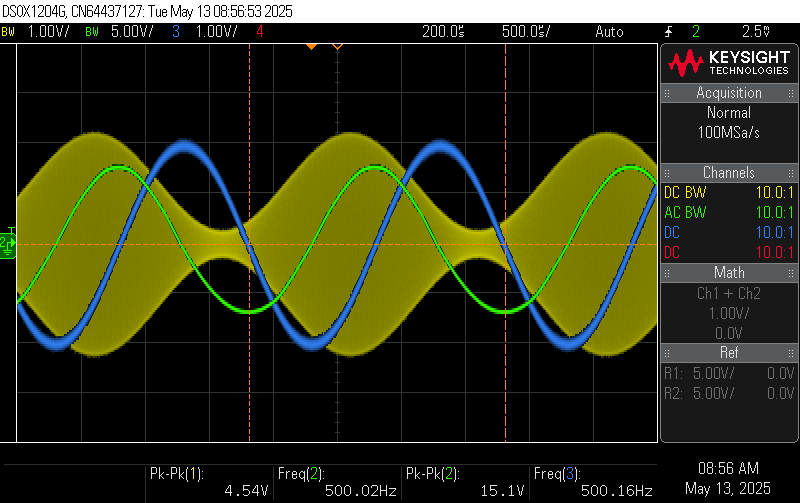}
    \caption{5 $V_{pp}$, 1 MHz, 80\% Mod Depth, 500 Hz}
    \label{fig:am_demod_fig_1}
\end{figure}

\begin{figure}[h]
    \centering
    \includegraphics[width=0.85\linewidth]{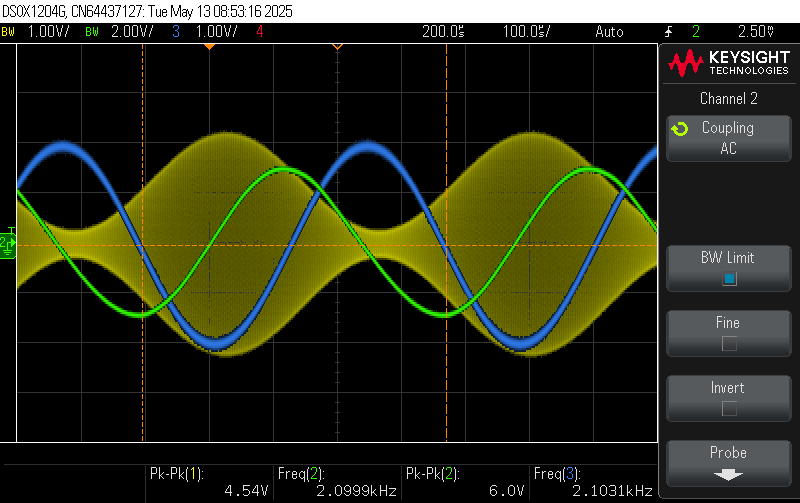}
    \caption{1.5 $V_{pp}$, 1 MHz, 80\% Mod Depth, 2.1 kHz}
    \label{fig:am_demod_fig_2}
\end{figure}

\begin{raggedright}
Next, we alter the modulation depth from 50\% to 100\% while holding the other parameters constant, observing again sinusoidal outputs with the expected frequency.
\newline
\end{raggedright}

\begin{figure}[htpb]
    \centering
    \includegraphics[width=0.85\linewidth]{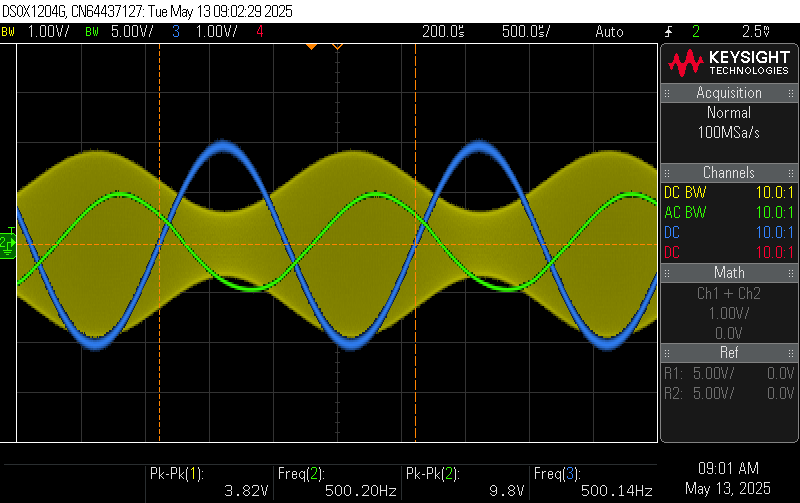}
    \caption{5 $V_{pp}$, 1 MHz, 50\% Mod Depth, 500 Hz}
    \label{fig:am_demod_fig_3}
\end{figure}

\begin{figure}[htpb]
    \centering
    \includegraphics[width=0.85\linewidth]{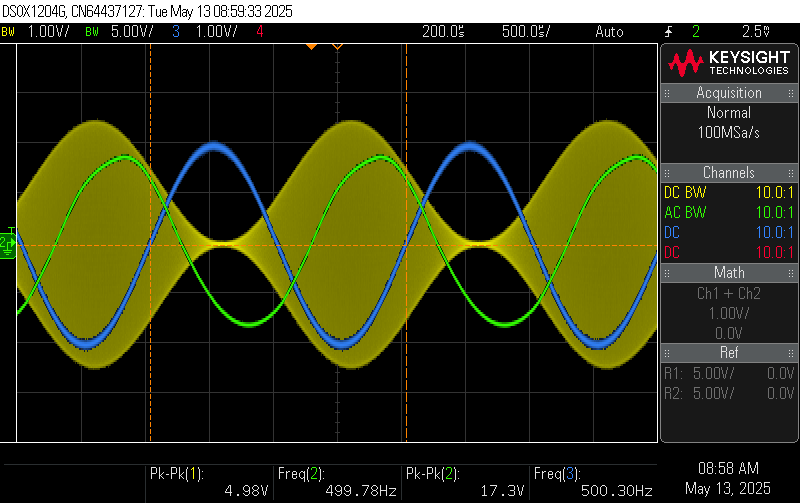}
    \caption{5 $V_{pp}$, 1 MHz, 100\% Mod Depth, 500 Hz}
    \label{fig:am_demod_fig_4}
\end{figure}

\begin{raggedright}
Finally, we sweep nearly the entire AM band to view how the demodulator performs at each end. It should be noted that although the performance degrades at 1.7 MHz with a nonsinusoidal output, it improves at higher frequencies.
\end{raggedright}

\begin{figure}[htpb]
    \centering
    \includegraphics[width=0.85\linewidth]{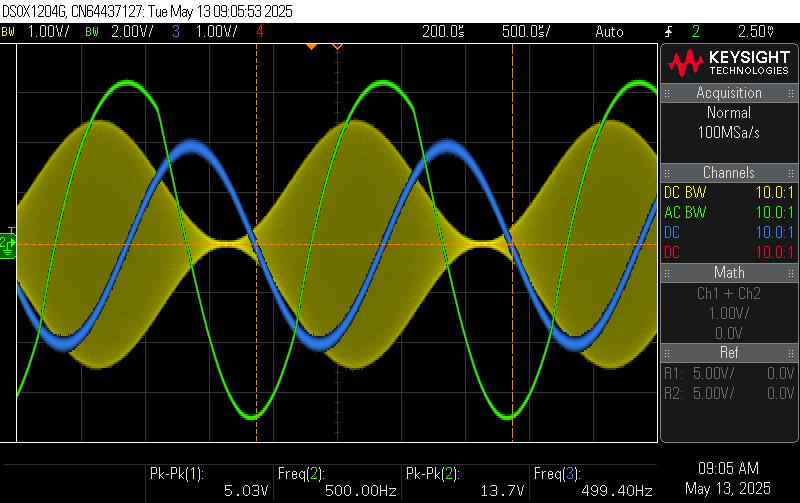}
    \caption{5 $V_{pp}$, 600 kHz, 100\% Mod Depth, 500 kHz}
    \label{fig:am_demod_fig_5}
\end{figure}

\begin{figure}[htpb]
    \centering
    \includegraphics[width=0.85\linewidth]{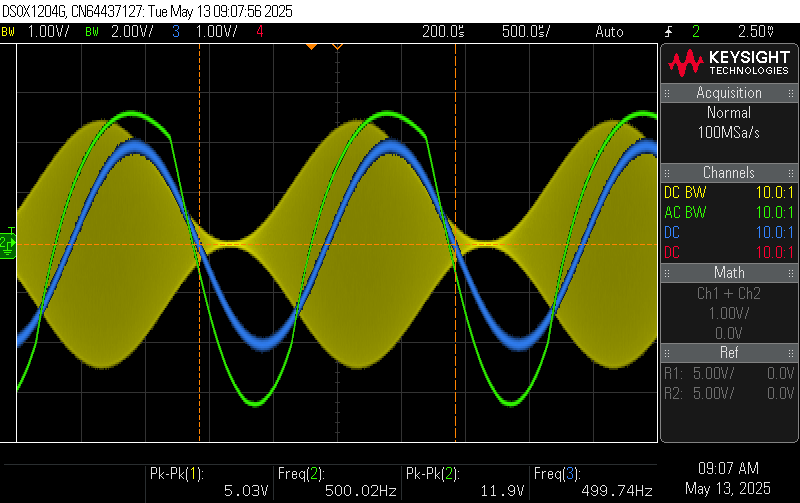}
    \caption{5 $V_{pp}$, 1.7 MHz, 100\% Mod Depth, 500 kHz}
    \label{fig:am_demod_fig_6}
\end{figure}

\subsubsection{Pre-Emphasis}
\begin{figure}[h!]
    \centering
    \includegraphics[width=0.9\linewidth]{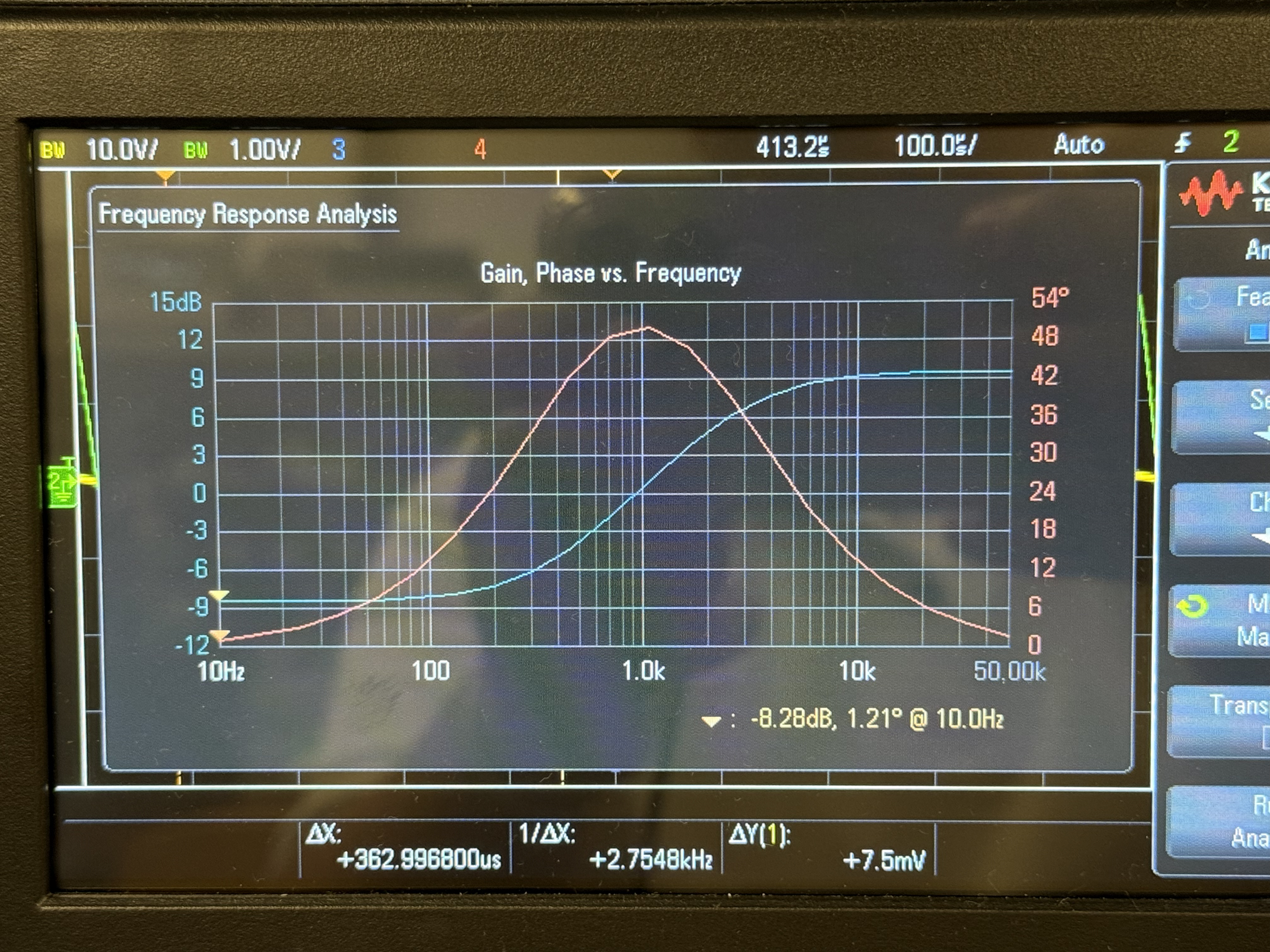}
    \caption{Scope measurements of the frequency response of the pre-emphasis circuit.}
    \label{fig:pre-emph}
\end{figure}

\begin{raggedright}
To evaluate the pre-emphasis circuitry, we wanted to validate that the components built on protoboard matched the expected frequency response from simulation. To verify these properties, we used the frequency response feature on the Agilent oscilloscopes in the lab, sweeping from 10 Hz to 50 kHz, with 20 points. As shown in the figure, the real-world frequency response of the pre-emphasis circuitry has a lower turning point at 300 Hz and an upper turning point at 2.5 kHz, exactly matching the simulation and confirming the behavior of the system.
\end{raggedright}



\subsubsection{Voltage Controlled Oscillator}

\begin{figure}[h!]
    \centering
    \includegraphics[width=0.5\linewidth]{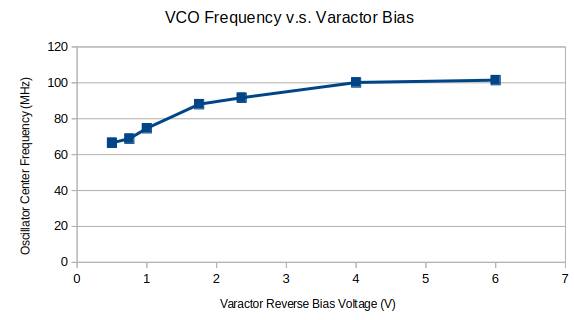}
    \caption{Frequency vs. Voltage for the VCO assembled on perfboard}
    \label{fig:v_vs_f}
\end{figure}
\begin{raggedright}
The constructed VCO met the criteria set out in section \ref{subsec:system_vco}. The center frequency is stable within $\pm 10 \, \mathrm{kHz}$, which does introduce some amount of noise into the output signal, but it is relatively small compared to the channel width of 200 kHz. Because the varactor diodes exhibit an approximately exponential change in capacitance due to a change in the reverse bias voltage \cite{BBY66datasheet}, output frequency also does not change linearly with bias voltage. As the bias voltage increases, the change in output frequency increases significantly as seen in ~\Cref{fig:v_vs_f}.
\newline
\newline
However, when an input signal is introduced to the system, the change in reverse bias voltage on the varactors is not in the volts, but rather in the millivolts. In that range, the behavior of the VCO is approximately linear, as the change in voltage is so small. As a result, this nonlinearity does not have a significant negative effect on the system as the VCO is still approximately linear around its center frequency.
\end{raggedright}

\subsection{Overall System Characterization}
\label{subsec:overall_sys}
\begin{raggedright}
    
The completed system operates over a frequency range of 66-102 MHz, which includes the least congested, and therefore most useful, portion of the FM broadcast band. By testing the system with a signal generator to produce an AM input for the system and a software defined radio to receive the output, we found that the audio quality from the system is more than passable. Qualitatively, middle and upper frequencies sound near-perfect, but low frequencies, especially those that are loud, tend to sound a bit distorted. However, the signal quality of the low frequencies did improve significantly after pre-emphasis, potentially suggesting that future changes would only need small changes to the frequency response of pre-emphasis to further improve sound quality. From a usability standpoint, the system would be difficult to tune for a non-technical user, but would only require small subsequent adjustments that are easier to make. For example, tuning became significantly easier with the addition of the ten-turn potentiometers.
\newline
\newline
To fully characterize the complete system, measurements of the output spectrum were made on the Agilent E4443A spectrum analyzer. 
\begin{figure}[h]
    \centering
    \includegraphics[width=0.5\linewidth]{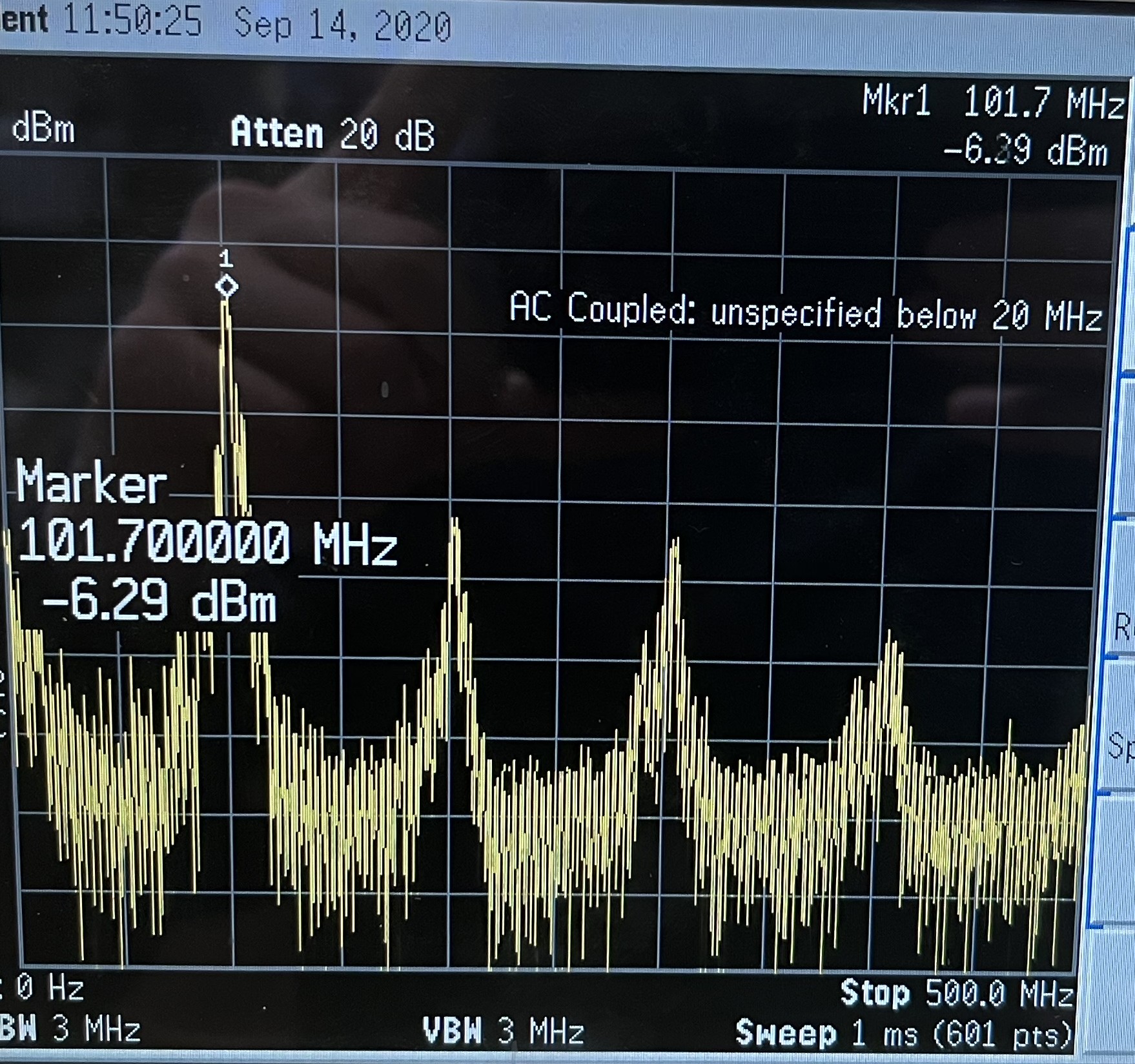}
    \caption{Harmonics of the oscillator present in the output spectrum.}
    \label{fig:harmonics}
\end{figure}

In Figure \ref{fig:harmonics}, the low spectral purity of the device can be observed. While the fundamental is about 15 dB stronger than the strongest harmonic, this could be further improved by adding a lowpass filter on the output. These harmonics are likely inherent to the function of the Colpitts oscillator. As mentioned in section \ref{subsubsec:colpitts_analysis}, the oscillator relies on the output voltage swing being large enough that the effective gain is reduced to 1, but the nonlinearities that are present in the transistor's large-signal operation will produce harmonics. 

\begin{figure}[H]
    \centering
    \includegraphics[width=0.5\linewidth]{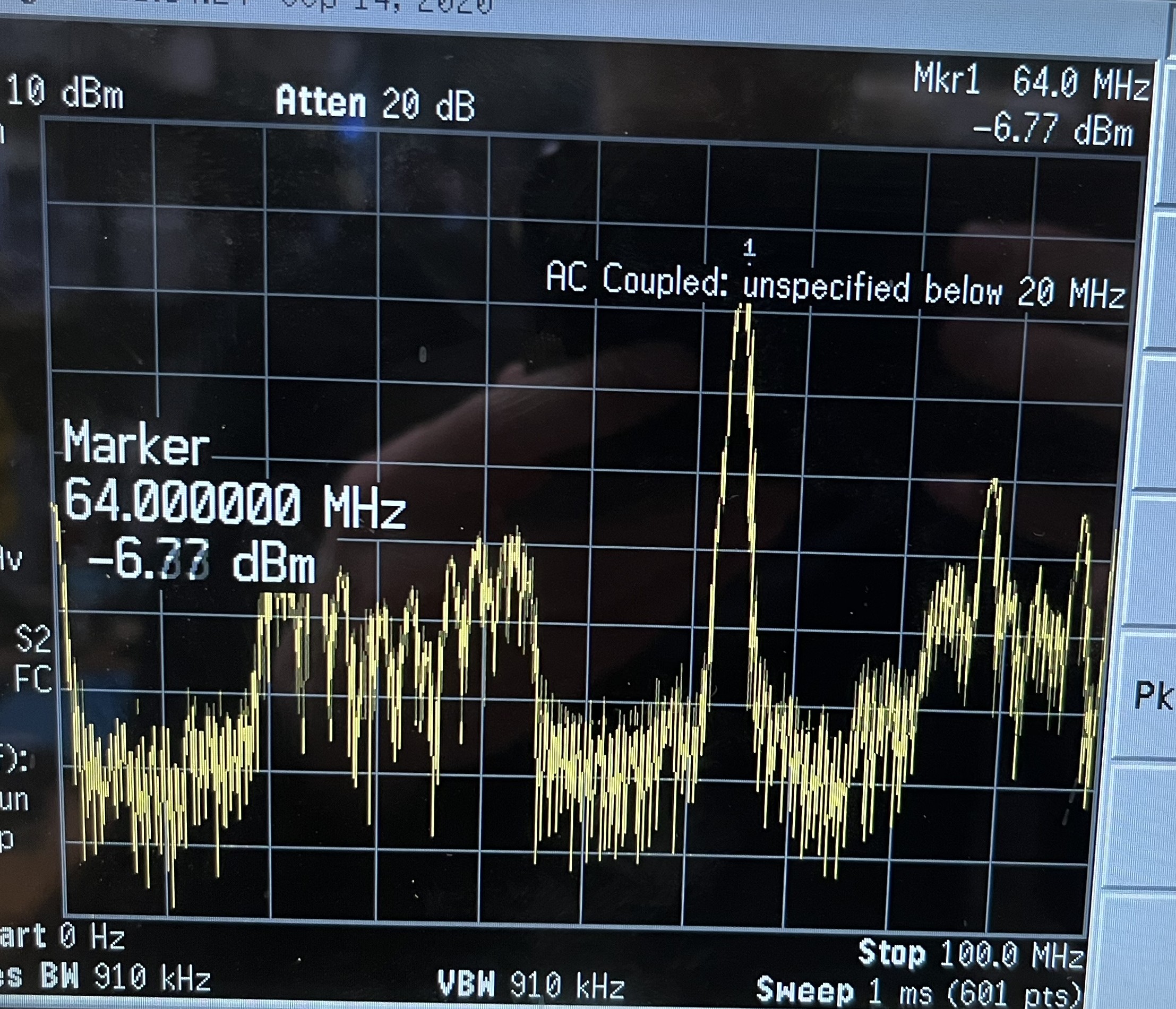}
    \caption{Intermodulation produces an image of the FM band at 20-40 MHz.}
    \label{fig:mixing}
\end{figure}

\end{raggedright}

\begin{raggedright}
An additional effect of the nonlinear devices in the system is intermodulation. This occurs when unintentional frequency mixing occurs, producing images of signals where they would not normally be expected. In figure \ref{fig:mixing}, there appear to be 3 separate signals. On the far right is the broadcast FM band, which is strong enough that it couples into our device. The signal to the left of that at 64 MHz is the center frequency of the oscillator. The signal even further to the left, from 20-40 MHz, is the image of the FM broadcast band that is created by the mixing product $f_{FM} - f_{LO}$.
\end{raggedright}

\section{Further Areas for Improvement}

There are a number of aspects where we strongly believe that our project can be extended. The first such section is based on our construction of oscillators and demodulators. While we use a full-wave rectifier-based demodulator, a standard industry implementation uses a superheterodyne receiver, which takes the varied AM band and converts it into an intermediate frequency, where a more accurate detector like the product detector can be used. In this direction, we would like to use a phase-locked loop based implementation where we can ensure that the input signal from crystal oscillator and the input AM signal are matched in phase, meaning that we can use mixers and other modules to perform multiplications on the signals. A similar approach can also be used to construct a PLL-based voltage-controlled oscillator.
\newline
\newline
Other directions that could be pursued include the construction of a PCB rather than perfboard for our project. The perfboard we used still had connected rails, which have built-in capacitance, but these effects could be better mitigated using shielded PCBs. The shielding would also potentially better protect our AM-to-FM converter from intermodulation from other FM signals, which was an issue for us in this project, even when putting the device in a metal box (which we attribute to the cabling). We believe that these future directions could take our project and elevate it even further, improving ease of use for the future.


\section{Conclusion}

At this project's conclusion, we have thoroughly demonstrated that we can take an AM signal, such as from a laptop's audio jack, and transform this into a FM signal that can be demodulated by an software-defined radio (SDR) and listened to once again with reasonable fidelity. During final testing, we demonstrated this for a range of different songs: Piano Man by Billy Joel, Free Bird from Lynyrd Skynyrd, and Hotel California by the Eagles. Additionally, we succeeded in integrating our project onto a bug-free perfboard, completing final testing with the perfboard. Throughout this project, we struggled to effectively tune the VCO so it was not co-located with an existing FM station. So, we added two ten-turn potentiometers, one to control the signal bandwidth, and another control the oscillation frequency. This significantly improved tuning and speaks to designing a system which is not just accessible to the original designers, but newcomers as well. Below is a photo of our final project with the discrete sections labeled:

\begin{figure}[h]
    \centering
    \includegraphics[width=\linewidth]{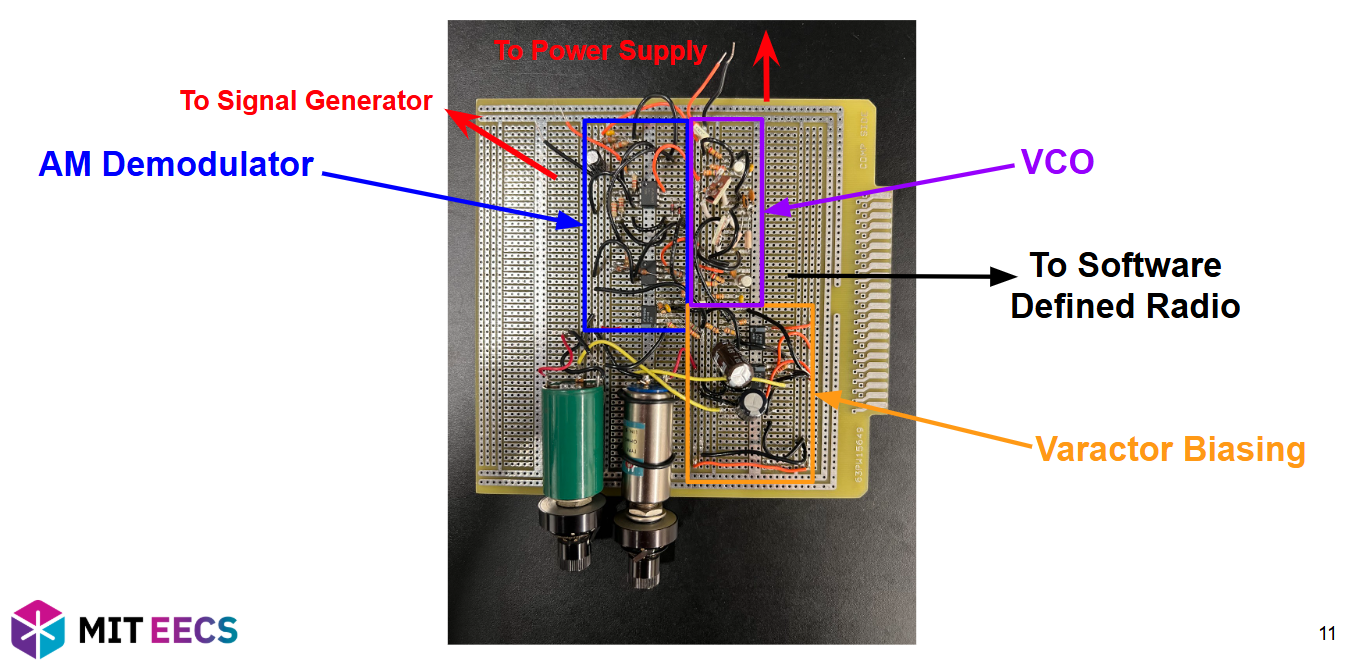}
    \caption{Final, Integrated Perf Board Project}
    \label{fig:final_product}
\end{figure}


\section{Acknowledgments}

A heartfelt thank you to everyone for their mentorship and support not just on this project but over the course of this term: Professor Coln, Tiffany, Dr. Larson, Dr. Carleton, Soojung, Anika, and Alec.

\pagebreak
\bibliographystyle{IEEEtran}
\bibliography{refs}

\end{document}